\documentclass[aps,prb,twocolumn,superscriptaddress]{revtex4-2}
\usepackage{times}
\usepackage{graphicx}
\usepackage{amsmath}
\usepackage{hyperref}
\usepackage{subcaption} 
\usepackage[percent]{overpic}
\usepackage{float}
\usepackage{xcolor}
\usepackage{ragged2e}

\begin{document} 

\title{Emergent Pair Density Wave and Incoherent Metallic State in a Strongly Correlated Doped System }

\author{Soham Maiti}
\affiliation{Department of Physics, Indian Institute of Technology Kharagpur, Kharagpur 721302, West Bengal, India}

\author{Nandan Pakhira}
\affiliation{Department of Physics, Kazi Nazrul University, Asansol, West Bengal 713340, India}

\author{A. Taraphder}
\affiliation{Department of Physics, Indian Institute of Technology Kharagpur, Kharagpur 721302, West Bengal, India}


\begin{abstract} 

We investigate the dynamical emergence of a state driven by the interplay between antiferromagnetism (AFM) and singlet $d$-wave superconductivity (SC) within a slave-rotor mean-field framework. By decomposing the electron into charge and spin degrees of freedom, the formalism captures strong-correlation effects beyond conventional mean-field approaches. A PDW order is found to emerge dynamically in the coexistence region of AFM and SC. The AFM–SC coexistence region is significantly modified with correlation, leading to a systematic shift of the tetra-critical point. The doping and temperature evolution of the AFM, SC, and PDW order parameters, together with the rotor condensate amplitude \(\phi\), which characterizes charge coherence, shows a crossover from a coherent to an incoherent metal with the suppression of coherent quasi-particle spectral weight. Moreover, in the AFM + (\(\phi \neq 0\)) region coherent quasi-particle bands coexist with incoherent Hubbard-like excitations, whereas only incoherent spectral features survive in the AFM + (\(\phi = 0\)) regime. The SC phase exhibits nodal quasi-particles consistent with \(d\)-wave pairing.
\end{abstract} 

\maketitle
\section{Introduction}

Strongly correlated electron systems display a remarkable interplay of competing and intertwined orders ~\cite{RevModPhys.87.457, Tu2016, Hayden2024}. Understanding how charge localization, magnetic ordering, and superconductivity influence each other remains a central challenge in condensed-matter physics. Considerable evidence suggests that several types of magnetic order, charge density wave (CDW) order, nematicity, and superconductivity—are not simply competing phases but components of a broader, intertwined-order landscape arising from strong electronic correlations. Despite significant theoretical and numerical progress, a unified understanding of how the various low-energy orders emerge and compete in strongly correlated systems continues to be lacking, owing in part to the absence of fully controlled analytical frameworks. \\

Among these intertwined orders, the pair-density-wave (PDW) \cite{PhysRev.135.A550, Larkin1965, PhysRevLett.88.117001, PhysRevLett.99.127003, Agterberg2008, Agterberg2020} state has attracted significant theoretical and experimental interest lately. A PDW order parameter is characterized by a gap function that varies periodically in the position space in a way that its spatial average vanishes. Theoretical studies have suggested that PDW order naturally emerges in strongly correlated systems, particularly in the proximity of stripes and other electronic orders \cite{PhysRevB.79.064515}. Earlier, numerical and variational studies of the doped \(t–J\) model \cite{Yang2009, PhysRevB.76.140505} hinted at pairing states with spatial modulation, while later analytical work proposed PDW state as a major element of the cuprate phase diagram \cite{PhysRevB.38.4596, PhysRevB.70.035114, PhysRevB.83.134515, PhysRevLett.62.2751, Axe1994, PhysRevLett.99.127003, PhysRevB.78.174529}. More recently, scanning tunneling microscopy has reported signatures consistent with PDW order inside vortex halos and stripe-ordered cuprates \cite{Hamidian2016}. Several microscopic mechanisms have been proposed to account for PDW formation, including Amperean pairing \cite{PhysRevX.4.031017} driven by spin-current fluctuations in RVB-type states \cite{PhysRevLett.98.067006} and competing superconducting instabilities near Pomeranchuk channels \cite{PhysRevB.89.165126}, as well as an analysis based on quasi-1D stripe models. Interestingly, a possible PDW ground state has been established in the strongly correlated one-dimensional Kondo–Heisenberg chain, providing theoretical evidence for finite-momentum pairing \cite{PhysRevLett.105.146403}. Collectively, these developments point to a scenario in which finite-momentum pairing arises from the underlying correlated electronic environment rather than being externally imposed. \\ 

Despite these progresses, the microscopic mechanism that enables an emergence of PDW order, remains debated. In particular, most of the existing theoretical approaches, ranging from strong-coupling \(t–J\) model studies to spin-fermion theories \cite{PhysRevB.49.4261}, treat electrons as quasi-particles with renormalized interactions, without explicitly accounting for a possible charge–spin separation. However, strong correlation physics is known to drastically modify quasi-particle coherence, superconducting order and magnetic correlations. This raises a natural question: Can the interplay between charge localization and magnetic order, by itself, drive the formation of a PDW state? Addressing this question requires a theoretical framework capable of treating charge fluctuations and spin ordering on equal footing. \\

The slave-rotor method \cite{PhysRevB.66.165111, PhysRevB.70.035114}, provides a useful route for such an investigation. In this approach, the electron operator is decomposed into a fermionic spinon that carries spin and a bosonic rotor that carries charge. The condensation of the rotor field, measured through the rotor condensate \(\phi = \langle e^{i\theta}\rangle\), controls the degree of quasi-particle coherence and encodes the strength of correlation effects. This framework has been successfully applied to study Mott transitions \cite{PhysRevB.83.134515} at half-filling, correlated metals \cite{PhysRevB.76.195101}, and unconventional superconductivity \cite{PhysRevB.75.245105}. Indeed, for strongly correlated systems, the static long-range orders and a possible spin-liquid state have been successfully described by this method. However, there is an alternative possibility of a dynamical generation of an order, albeit in the presence of other static non-zero orders, suggested earlier\cite{Psaltakis1983, PhysRevB.62.9083, Maitra2001}. To the best of our knowledge, such dynamical generation of PDW order within the slave-rotor formulation has not been studied so far. \\ 

In the late 1990s, Zhang and co-workers introduced the SO(5) theory as an attempt at a unifying description for antiferromagnetism and \(d\)-wave superconductivity \cite{Zhang1997}. Within the mean-field framework, the coexistence of antiferromagnetism and a singlet superconducting order of a given symmetry dynamically induces a \(\pi\)-triplet pairing \cite{PhysRevB.62.9083, Maitra2001}. The alternate superconducting state can be viewed as a pair-density-wave (PDW) order closely related to \(\pi\)-triplet pairing \cite{PhysRevLett.105.146403, PhysRevB.89.165126}. In this work, we investigate the dynamical emergence of PDW superconductivity in a model that incorporates AFM order, singlet pairing, and charge fluctuations treated self-consistently within the slave-rotor framework. We demonstrate that the combined effects of magnetic order and rotor-controlled charge coherence can naturally generate a finite-momentum pairing amplitude even when no explicit PDW interaction is introduced in the Hamiltonian. By analyzing the temperature–doping phase diagram, we show that the PDW state appears only within the region where the AFM and \(d\)-wave singlet SC coexist, and its stability is controlled by the magnitude of the driving potential of the AFM and SC. \\ 

Strongly correlated electron systems often exhibit a breakdown of conventional quasi-particle descriptions, leading to a crossover between coherent and incoherent electronic excitations \cite{PhysRevB.70.035114, Georges1996, PhysRevB.91.075124}. In this context, the distinction between a coherent metal with well-defined quasi-particles and an incoherent metal characterized by the absence of long-lived excitations has attracted considerable attention. In the conventional slave–rotor studies, incoherent metallic behavior and incoherent-metal phases are typically discussed in close proximity to the Mott insulating state at or near half filling, where the loss of quasi-particle coherence is driven by strong correlation effects associated with the Mott transition \cite{RevModPhys.68.13}. In contrast, the present work focuses on finite doping and finite temperature regimes, where no Mott insulating phase is realized in square lattices. Remarkably, we find that a coherent–incoherent crossover and the emergence of an incoherent-metal state arise even away from half-filling. Within this framework, the loss of charge coherence is governed by the thermal suppression of the rotor condensate rather than proximity to a Mott transition. \\ 

Our results demonstrate that strong-correlation effects, encoded through the slave–rotor condensate, provide a natural mechanism for the emergence of PDW order from the interplay between AFM and SC. Beyond the formation of modulated pairing, the framework reveals a unified description of electronic excitations in correlated systems, characterized by a coherent–incoherent crossover governed by the rotor condensate. In particular, the phase diagram spans coherent and incoherent metallic regimes, and the AFM phase further separates into two distinct sectors delineated by charge coherence, resulting in markedly different low-energy spectral properties despite similar underlying magnetic order. These results highlight how charge–spin separation and the interplay of competing orders shape both the ordering phenomena and the spectral properties, providing a broader perspective on PDW physics in strongly correlated systems.\\

The rest of the paper is organised as follows: in Section II we introduce the model Hamiltonian and the formalism. In Section III we study the phase diagram and the emergence of the PDW state. In Section IV we study the spectral features of the model. Finally, in Section V, we conclude.

\section{Model and Formalism}
To investigate the competition and coexistence of intertwined electronic orders in strongly correlated systems, we employ a phenomenological slave rotor mean-field approach on a two-dimensional square lattice, treating anti-ferromagnetism (AFM), singlet superconductivity (SC), and triplet pair-density wave (PDW) order  on equal footing. Within this framework, we examine the emergence and interplay of the AFM, singlet SC, and triplet PDW phases. \\
\begin{equation}
\begin{aligned}
H =& -t\sum_{\langle i,j \rangle ,\sigma}
(c_{i,\sigma}^{\dagger}c_{j,\sigma} + h.c.)
+ U\sum_{i} n_{i,\uparrow}n_{i,\downarrow} \\
& - W_{s}\sum_{i}(C_{i}^{\dagger}\langle C_{i} \rangle + h.c.)
- W_{p}\sum_{i}(D_{i}^{\dagger}\langle D_{i}\rangle + h.c.) 
\end{aligned}
\end{equation}
\noindent \(\langle i,j \rangle\) is the nearest neighbour interaction. \(t\) is a hopping amplitude between nearest neighbour sites. \(U\) is the onsite repulsive term. \(W_{s}, W_{p}\) are the driving potentials for SC, PDW states, respectively. \(C_{i}, D_{i}\) are defined by 
\begin{subequations}
\begin{align}
C_{i} &= \sum_{\eta}\zeta(\eta)
\bigl(c_{i,\uparrow}c_{i+\eta,\downarrow}
- c_{i,\downarrow}c_{i+\eta,\uparrow}\bigr) \\
D_{i} &= \sum_{\eta}\zeta(\eta)
\bigl(c_{i,\uparrow}c_{i+\eta,\downarrow}
+ c_{i,\downarrow}c_{i+\eta,\uparrow}\bigr) 
\end{align}
\end{subequations}
\noindent Here, \(\zeta(\eta)\) are the form factors which are defined by
\(\zeta(\eta) = +1\) when \(\eta = (\pm 1, 0)\) , \(\zeta(\eta) = -1\) when \(\eta = (0, \pm 1)\). We choose the form factors \(\zeta(\eta)\) to impose the \(d\)-wave symmetry properties on the singlet superconducting (SC) order parameter and triplet pair density wave (PDW) order parameter. 
To treat the on-site interaction within the slave-rotor framework, the Hubbard term is decomposed into its charge and spin contributions. Following usual practice, we decouple the interaction into spin and charge channels. The former accounts for antiferromagnetic correlations and enters the spinon Hamiltonian, while the latter governs charge fluctuations in the rotor sector. 
 
In the slave rotor representation, electron creation (annihilation) operator \(c_{i,\sigma}^{\dagger} (c_{i,\sigma})\) is expressed as \(c_{i,\sigma} = f_{i,\sigma}e^{-i\theta_{i}}, \:\:\: c_{i,\sigma}^{\dagger} = f_{i,\sigma}^{\dagger}e^{i\theta_{i}}\) \cite{PhysRevB.66.165111}. \(f_{i,\sigma}^{\dagger} (f_{i,\sigma})\) denotes the spinon creation (annihilation) operator, while \(e^{i\theta_{i}} (e^{-i\theta_{i}})\) represents the rotor, or chargon, creation (annihilation) operator. A local constraint relates the average chargon and spinon densities,\, \(\langle n_{i}^{\theta} \rangle \) and \(\langle n_{i}^{f} \rangle \), respectively \cite{PhysRevB.76.195101}:\,
\(\langle n_{i}^{\theta} \rangle + \langle n_{i,\uparrow}^{f}\rangle + \langle n_{i, \downarrow}^{f} \rangle = 1.\) Expressing the electron creation (annihilation) operators in terms of spinon creation (annihilation) operators, the Hamiltonian is decoupled into spinon and rotor sectors. The electronic ground state takes the form \(|\psi \rangle  = |\psi\rangle_{f}|\psi \rangle_{\theta}\) where \(|\psi \rangle_{f}\) and \(|\psi \rangle_{\theta}\) are the spinon and rotor components of the wavefunction, respectively. On relaxing the local constraints to a global one (using a spinon and a rotor chemical potential) in the homogeneous theory, the resulting spinon and rotor Hamiltonians are: \\
\begin{equation}
\begin{aligned}
H_{f} =& -t \sum\limits_{\langle i,j \rangle \sigma}(f_{i,\sigma}^{\dagger}f_{j,\sigma} B_{i,j} + h.c.) - \\
&W_{m} \sum\limits_{i}^{}(M_{f,i}^{\dagger} \langle M_{f,i} \rangle + h.c) - W_{s} \sum\limits_{i} (C_{f,i}^{\dagger} \langle C_{f,i} \rangle +  \\
&h.c. ) -  W_{p}\sum\limits_{i} (D_{f,i}^{\dagger} \langle D_{f,i} \rangle + h.c.) - \mu_{f}\sum\limits_{i,\sigma}n_{i,\sigma}^{f} \\
\end{aligned}
\end{equation}
\begin{equation}
H_{\theta} = -2t\sum\limits_{i,j}^{}\chi_{i,j}(e^{i\theta_{i}}e^{-i\theta_{j}} + h.c) + \frac{U}{4}\sum\limits_{i}^{}(n_{i}^{\theta})^{2} - \mu_{\theta}\sum\limits_{i}n_{i}^{\theta} 
\end{equation}
\noindent where, \(W_m = \frac{U}{4}\), \(B_{i,j} = \langle \psi_{\theta}|e^{i\theta_{i}}e^{-i\theta_{j}}|\psi_{\theta}\rangle, \chi_{i,j} = \langle \psi_{f}|f_{i,\sigma}^{\dagger}f_{j,\sigma}|\psi_{f}\rangle\).
\(\mu_{f}, \mu_{\theta}\) are the spinon and rotor chemical potentials. \\ 

Starting from the spinon Hamiltonian in Eq.(3), a mean-field decoupling of all terms in the Hamiltonian yields the parameters \(M_{f,i}, C_{f,i}, D_{f,i}\) defined by 
\begin{subequations}
\begin{align}
M_{f,i} &= (n_{i, \uparrow}^{f} - n_{i,\downarrow}^{f}) \\
C_{f,i} &=  \sum\limits_{\eta}^{}\zeta(\eta)(f_{i,\uparrow}f_{i+\eta,\downarrow} - f_{i,\downarrow}f_{i+\eta,\uparrow})\\
D_{f,i} &=  \sum\limits_{\eta}^{}\zeta(\eta)(f_{i,\uparrow}f_{i+\eta,\downarrow} + f_{i,\downarrow}f_{i+\eta,\uparrow})  
\end{align}
\end{subequations}
\vspace{2 mm}
\noindent The AFM order parameter, spinon SC gap parameter and spinon PDW order parameter denoted by \(\Delta_m\), \(\Delta_{sf}\) and \(\Delta_{pf}\) respectively, are given by 
\vspace{2 mm}
\begin{subequations}
\begin{align}
\Delta_{m}\cos(Q.r_{i}) &= \langle n_{i,\uparrow}^{f} - n_{i,\downarrow}^{f}\rangle = \langle f_{i, \uparrow}^{\dagger}f_{i,\uparrow} - f_{i,\downarrow}^{\dagger}f_{i,\downarrow}\rangle \\
\Delta_{sf} &= \zeta(\eta)\langle f_{i,\uparrow}f_{i+\eta,\downarrow} - f_{i,\downarrow}f_{i+\eta,\uparrow} \rangle \\ 
\Delta_{pf}\cos(Q.r_{i}) &= \zeta(\eta)\langle f_{i,\uparrow}f_{i+\eta,\downarrow} + f_{i,\downarrow}f_{i+\eta,\uparrow} \rangle 
\end{align}
\end{subequations}

\noindent where \(Q=(\pi,\pi)\) is the commensurate ordering wave vector in a square lattice. The spinon Hamiltonian in the momentum space is 

\begin{equation}
\begin{aligned}
H_{k} =& \sum\limits_{k,\sigma}\epsilon_{k}(f_{k,\sigma}^{\dagger}f_{k,\sigma} + h.c.) -  \\
& W_{m}\Delta_m\sum\limits_{k}\Big[(f_{k,\uparrow}^{\dagger}f_{k + Q,\uparrow} - f_{k,\downarrow}^{\dagger}f_{k + Q,\downarrow}) + h.c.\Big] -   \\
& W_{s}\Delta_{sf}\sum\limits_{k}^{}g(k)(f_{k,\uparrow}f_{-k,\downarrow} + h.c.) - \\
& W_{p}\Delta_{pf}\sum\limits_{k}g(k)\Big[(f_{k,\uparrow}f_{-k-Q,\downarrow} + f_{-k,\downarrow}f_{k+Q,\uparrow}) + h.c.\Big]  \\
\end{aligned}
\end{equation}

\noindent where, \(\epsilon_{k} = -2tB(\cos k_{x} + \cos k_{y}) - \mu_{f}, \:\: g(k) = \cos k_x - \cos k_y \).

\vspace{2 mm}
This Hamiltonian is diagonalized to yield the energy dispersion relations \(\pm E_{\alpha = \pm }\). From the resulting partition function \(Z_{0}\), the free-energy functional is calculated as  
\begin{equation}
F = F_{0} + \langle H - H_{MF} \rangle_{MF} 
\end{equation}
\noindent where, \(F_{0} = -kT\log{Z_{0}}\). The second term represents the expectation value with respect to the reference mean-field system and accounts for the constant energy contribution arising from the mean-field approximation. The self-consistent equations are obtained as usual by extremizing the free energy functional: \(\partial F/\partial \Delta_{m} =\partial F/\partial \Delta_{sf} = \partial F/\partial \Delta_{pf} =0\). The chemical potential \(\mu_f\) is determined through the relation \(n^f= -\partial F/\partial \mu_f\). The resulting self-consistent equations are  

\begin{equation}
\begin{aligned}
\Delta_m =& \frac{1}{2N}\sum\limits_{k}\sum\limits_{\alpha = \pm} \Big [\gamma_m  + \frac{\alpha}{2f} (\epsilon_k + \epsilon_{k+Q})^{2}(\gamma_m + \gamma_s\gamma_p)\Big]\\
&\frac{1}{E_{\alpha}}\tanh\Big(\frac{\beta E_{\alpha}}{2}\Big)   \\
\end{aligned}
\end{equation}

\begin{equation}
\begin{aligned}
\Delta_{sf} =& \frac{1}{2N}\sum\limits_{k}\sum\limits_{\alpha = \pm } \Big[\gamma_s g(k) +\frac{\alpha}{2f}\big(\gamma_m \gamma_p (\epsilon_k + \epsilon_{k+Q})+ 2\gamma_s\gamma_p^{2}\big)\Big]\\
&\frac{1}{E_{\alpha}}\tanh\Big(\frac{\beta E_{\alpha}}{2}\Big) \\
\end{aligned}
\end{equation}

\begin{equation}
\begin{aligned}
\Delta_{pf} =& \frac{1}{2N}\sum\limits_{k}\sum\limits_{\alpha = \pm } \Big[\gamma_p g(k) + \frac{\alpha}{2f}\big(\gamma_p(\epsilon_k - \epsilon_{k+Q})^{2} + 2(\gamma_m\gamma_s \\  
&(\epsilon_k + \epsilon_{k+Q}) + 4\gamma_s^{2}\gamma_p)\big)\Big]\frac{1}{E_{\alpha}}\tanh\Big(\frac{\beta E_{\alpha}}{2}\Big) \\
\end{aligned}
\end{equation}

\vspace{0.5 cm}
\noindent Here, \(W_m\Delta_m = \gamma_m, W_s\Delta_{sf} g(k) = \gamma_s, W_p\Delta_{pf} g(k) = \gamma_p\) and \(f\) is determined from the expression for the quasi-particle energy eigenvalues, 
\begin{equation}
\begin{aligned}
E_{\alpha = \pm} =& \Big[(\epsilon_k^{2} + \epsilon_{k+Q}^{2})/2 + \sum\limits_{i=m,s,p} (W_{i}\Delta_{i} \times \\
&\Xi_{i})^{2} \pm f\Big]^{2} 
\end{aligned}
\end{equation}
\noindent where $\Xi_i$ are the appropriate symmetry factors (e.g., for AFM $i=m, \, \Xi_{i}=1$, for SC and PDW $i=s,\, p, \,\Xi_{i}=g(k)$).  
\vspace{0.3 cm}

\textbf{Two-Site Mean Field Theory:\,}
In this case, we solve the rotor Hamiltonian using the self-consistent cluster mean-field theory, which offers a significant improvement over the conventional single-site approximation. The key idea behind the cluster mean-field approach is to identify a finite cluster of lattice sites that captures essential local correlations, while the interaction of this cluster with the rest of the lattice is treated at the mean-field level. Such a cluster, embedded in an effective field, includes inter-site correlations that are neglected in single-site description. The resulting rotor Hamiltonian is 
\begin{equation}
\begin{aligned}
H_{\theta} =& -2t\chi(e^{i\theta_{1}}e^{-i\theta_{2}} + h.c.) - 6t\chi \phi (e^{i\theta_{1}} + e^{i \theta_{2}} + h.c.) + \\
&\frac{U}{4} (n_{1}^{\theta})^{2} + \frac{U}{4}(n_{2}^{\theta})^{2} - \mu_{\theta}(n_{1}^{\theta} + n_{2}^{\theta}) 
\end{aligned}
\end{equation}
The rotor Hamiltonian matrix is constructed in \(|n_{1}^{\theta}, n_{2}^{\theta}\rangle\) basis and diagonalized to obtain the full set of eigenvalues and eigenstates. The thermal averages of rotor condensate \(\phi\), the rotor kinetic energy \(B\) and doping \(\delta\) are then evaluated at different temperatures. The values of \(\phi\), \(B\) are obtained self-consistently and doping \(\delta\) is controlled by \(\mu_\theta\). The resulting rotor kinetic energies \(B_1 = \phi^{2}\) for next-nearest neighbour bonds and the corresponding \(B\) for nearest neighbours serve as inputs to the spinon Hamiltonian.

\section{Competing Orders and the Emergence of the Pair-Density-Wave State}
The resulting phase diagram includes coexisting antiferromagnetic (AFM), superconducting (SC), and pair-density wave (PDW) orders, showing how correlation affects the phase landscape. Phase diagrams from single-site and two-site frameworks are qualitatively similar, with minor shifts in the phase boundaries. 
 
\subsection{Dynamic Generation of Pair-Density Wave and Emergence of Incoherent Metal region}

\begin{figure}[H]
    \centering
    \includegraphics[width=1.0\linewidth]{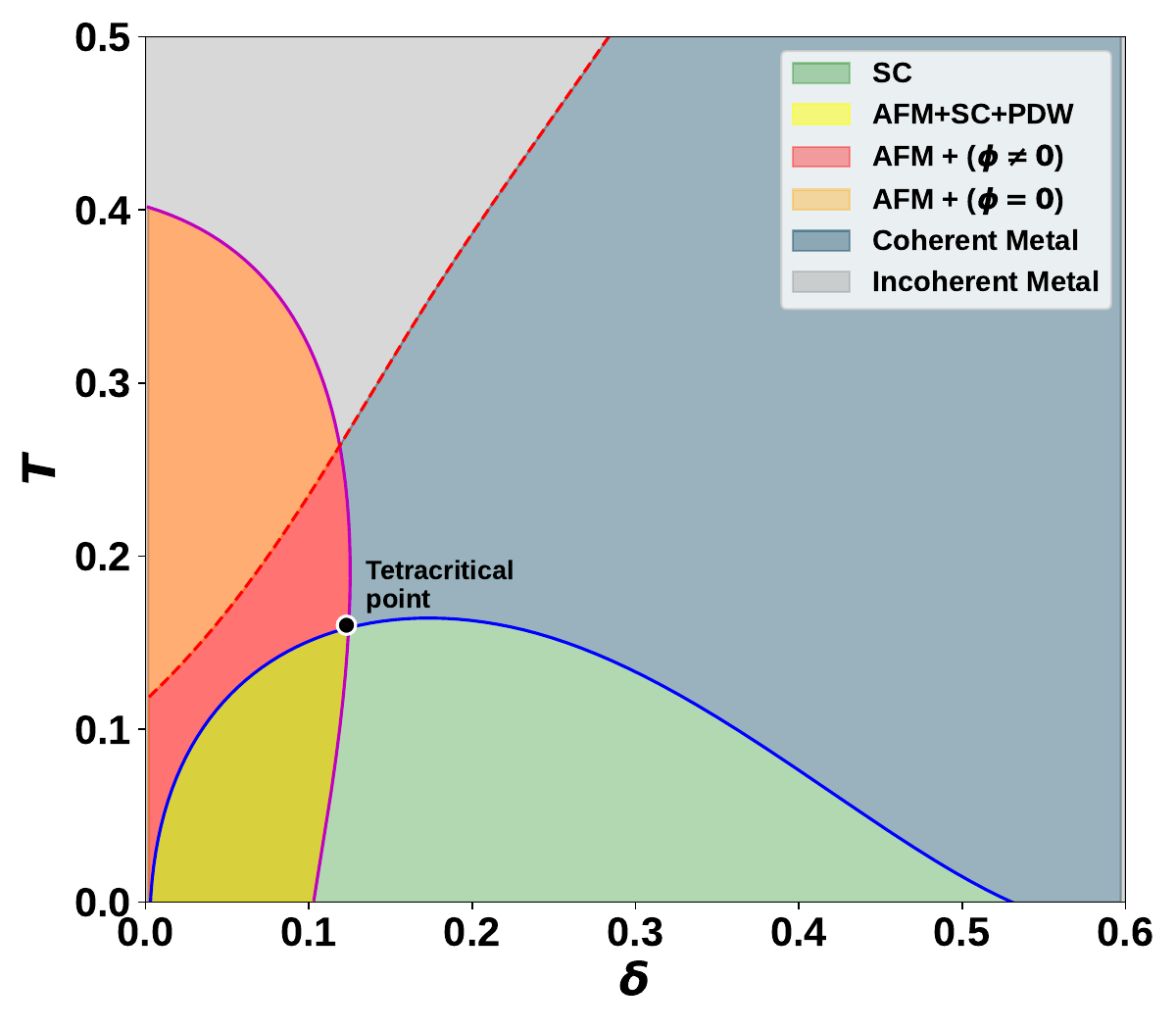}
    \caption{\justifying Temperature–doping (\(T - \delta\)) phase diagram for the dynamical emergence of a \(d\)-wave triplet pair-density-wave (PDW) from the coexistence of antiferromagnetism (AFM) and \(d\)-wave singlet superconductivity (SC) and the emergence of an incoherent metal region.}
    \label{fig:placeholder}
\end{figure} 

In Fig. 1, is shown the spontaneous emergence of \(d\)-wave triplet pair-density wave (PDW) state. The interaction strengths are set as \(W_{m} = 1.5, W_{s} = 1.0, W_{p} = 0\) and \(U = 6\). While mean-field theory already allows for the dynamical generation of a \(\pi\)-triplet (PDW-like) pairing channel in the presence of AFM and SC state\cite{PhysRevB.62.9083, Maitra2001}, the correlation effects stabilize the coexistence of AFM and singlet SC. In this regime, the \(d\)-wave triplet PDW order emerges dynamically despite the absence of an explicit driving potential. A notable feature of the phase diagram is the tetra-critical point \cite{PhysRevLett.32.1350}, where the AFM, SC, AFM+SC+PDW, and normal metallic phases intersect. This multi-critical point signifies the competition between magnetic, superconducting, PDW orders and separates four distinct thermodynamic phases. The phase boundaries are determined by solving the coupled spinon–rotor Hamiltonian self-consistently and locating the points at which the corresponding order parameters vanish. The superconducting order parameter takes the form \(\Delta_{sf}\phi^{2}\), indicating that both the spinon pairing amplitude \(\Delta_{sf}\) and the rotor condensate \(\phi\) must be nonzero for superconductivity to appear. Thus, the non-SC region is characterized by \(\Delta_{sf} = 0\), rather than by the vanishing of \(\phi\). An analogous criterion applies to the PDW order as well because it is defined by \(\Delta_{pf}\phi^{2}\). In contrast, the AFM order parameter \(\Delta_{m}\), does not factorize into spinon and rotor components; therefore, a finite \(\Delta_m\) implies the presence of AFM order. In Fig. 1, we identify two additional regions in the \(T-\delta\) plane: a coherent metallic phase characterized by a finite rotor condensate (\(\phi \neq 0\)), and an incoherent metallic phase where the rotor condensate vanishes (\(\phi = 0\)). We further observe regions where AFM order coexists with both coherent and incoherent metallic behavior. This coexistence arises naturally within the slave-rotor framework, as AFM order is governed by the spinon sector only, whereas \((\phi \neq 0)\) encodes charge coherence. Consequently, AFM order persists in both coherent and incoherent regimes.

When \(\phi \neq 0\), the charge excitations have well-defined quasi-particles. In contrast, when \(\phi = 0\), the system enters into an incoherent metallic regime. Importantly, within the two-site cluster formulation, \(\phi=0\) does not imply \(B=0\), since \(B = \langle e^{i \theta_{1}}e^{-i \theta_{2}} \rangle \neq \phi^{2}\). The finite value of \(B\) therefore sustains a nonzero spinon hopping amplitude. By comparison, in the single-site formulation, \(\phi = 0\) enforces \(B = \phi^{2} = 0\), collapsing the spinon dispersion to \(\epsilon_k = -\mu_f\). In the single-site SRMFT, the AFM + \((\phi = 0)\) phase corresponds to a fully localized spin sector, with vanishing spinon hopping. In contrast, within the two-site formalism, the AFM  (with \(\phi = 0)\) phase retains a finite spinon hopping, giving rise to dispersive spin excitations and short-range spatial correlations, even though the charge sector remains incoherent (\(\phi = 0\)).

\subsection{Evolution of the AFM, SC, PDW Phase with Interaction Strength \(U\)} 
Fig. - 2(a) and 2(b) illustrate the evolution of the AFM and SC phase boundaries in the \(T-\delta\) plane as a function of the interaction strength \(U\). The corresponding evolution of the PDW coexistence region for several representative values of \(U\) is presented in Fig. 2(c), while Fig. 2(d) provides a three-dimensional visualization of the PDW coexistence surface in the (\(\delta,T,U\)) parameter space. For the rest of the paper, we set $t=1$ and measure all parameters in units of $t$. Throughout this paper, \(U\) is varied in the range \(6 \leq U \leq 10\). We restrict our analysis to this range because, for smaller values of \(U\), the slave-rotor mean field theory is less applicable in the weakly correlated regime.  For these calculations, we set \(W_s = 1.0\) and \(W_p = 0\).

In Fig. 2(a), we observe that the phase boundary of AFM expands with increasing \(U\). This behavior arises because the effective driving potential for AFM order acquires a direct contribution from the interaction strength \(U\). Consequently, stronger onsite repulsion stabilizes the AFM phase and enlarges its domain in the phase diagram. Fig. 2(b) shows the evolution of SC phase boundary with the variation of \(U\). Where the AFM region increases with increase of \(U\), at the same time, the SC region decreases with increase of \(U\). The reduction of the SC region with increasing correlation strength \(U\) arises from two primary effects: enhanced electron localization and direct competition with AFM. As \(U\) increases, charge fluctuations are suppressed, reducing electronic mobility and thereby weakening the formation of Cooper pairs. Simultaneously, strengthened AFM order reconstructs the low-energy electronic structure and opens up gap over large portions of the Fermi surface, further suppressing superconducting correlations. The combined effect of reduced carrier mobility and AFM–SC competition leads to the systematic shrinkage of the SC dome at larger \(U\).

\begin{figure}[htbp]
\centering

\begin{subfigure}[t]{0.222\textwidth}
    \centering
    \begin{overpic}[width=\linewidth]{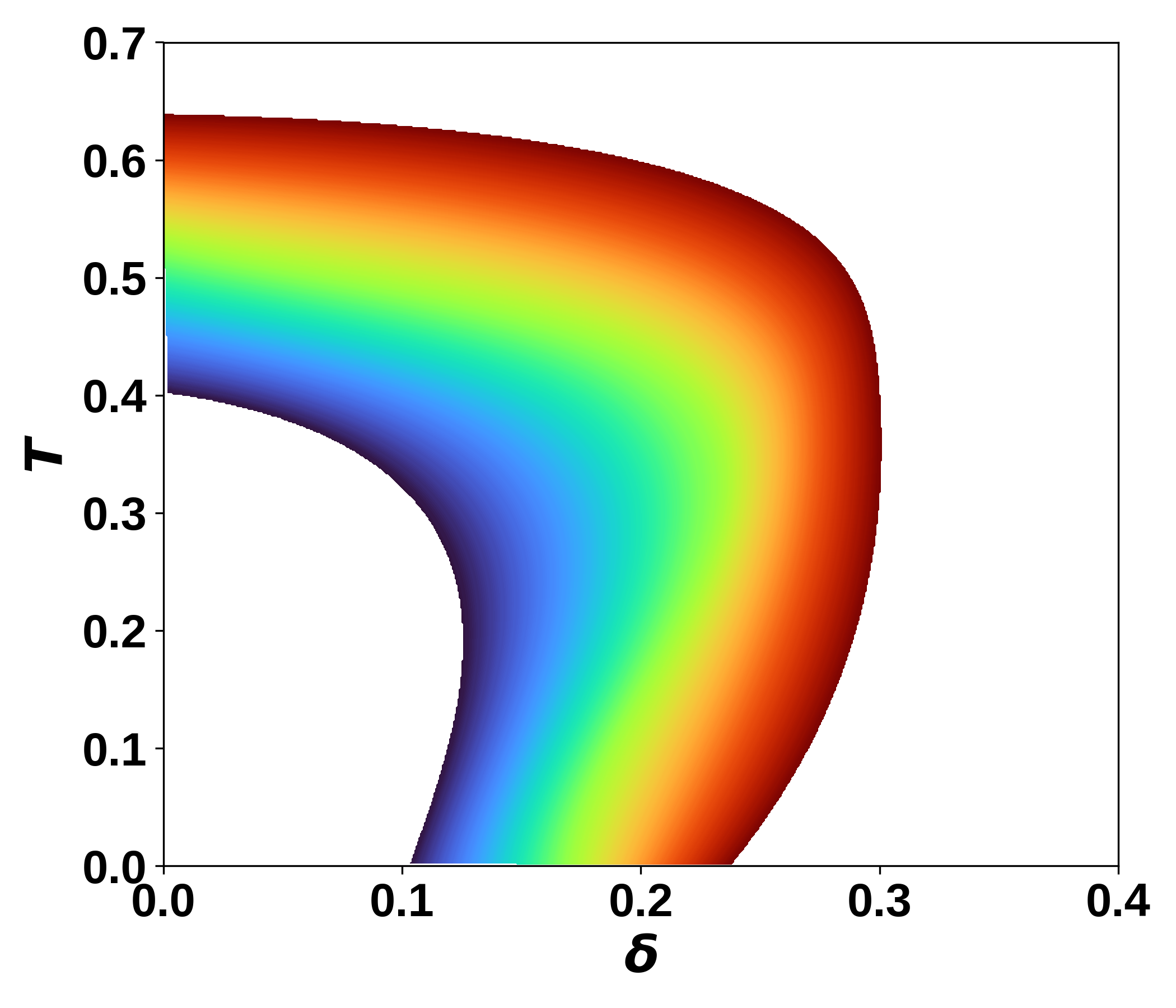}
        \put(75,70){\color{red}\small\bfseries (a)}
    \end{overpic}
\end{subfigure}
\hfill
\begin{subfigure}[t]{0.222\textwidth}
    \centering
    \begin{overpic}[width=\linewidth]{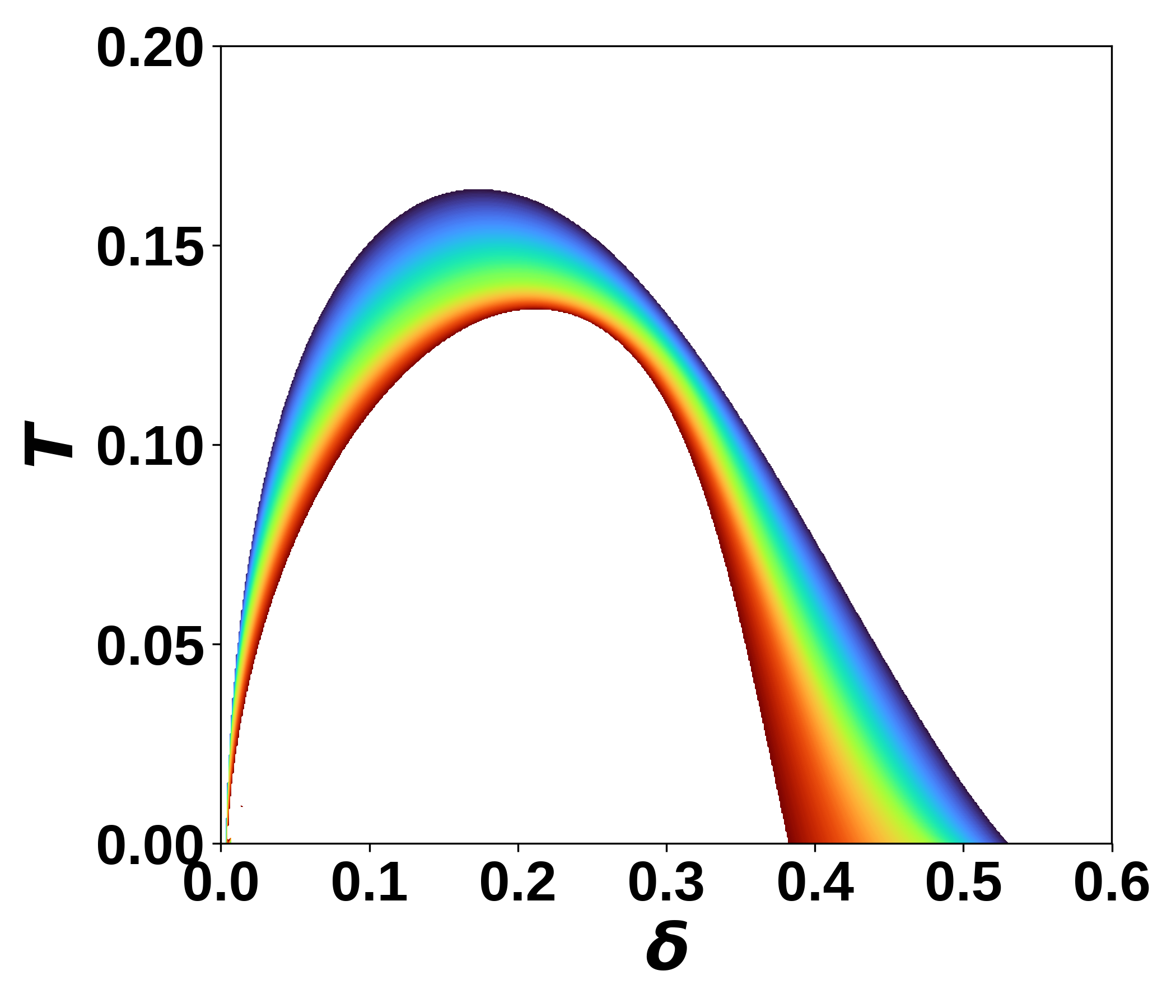}
        \put(75,70){\color{red}\small\bfseries (b)}
    \end{overpic}
\end{subfigure}
\hfill
\begin{subfigure}[t]{0.027\textwidth}
    \centering
    \includegraphics[width=\linewidth,height=3.5cm]{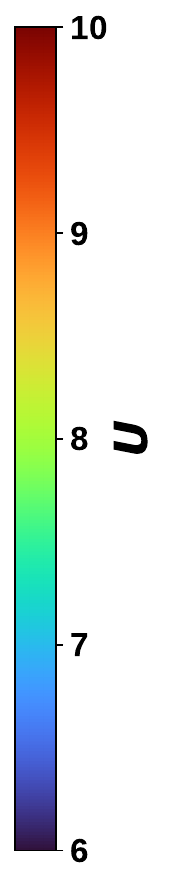}
\end{subfigure}

\vspace{0.4cm}


\begin{subfigure}[t]{0.222\textwidth}
    \centering
    \begin{overpic}[width=\linewidth]{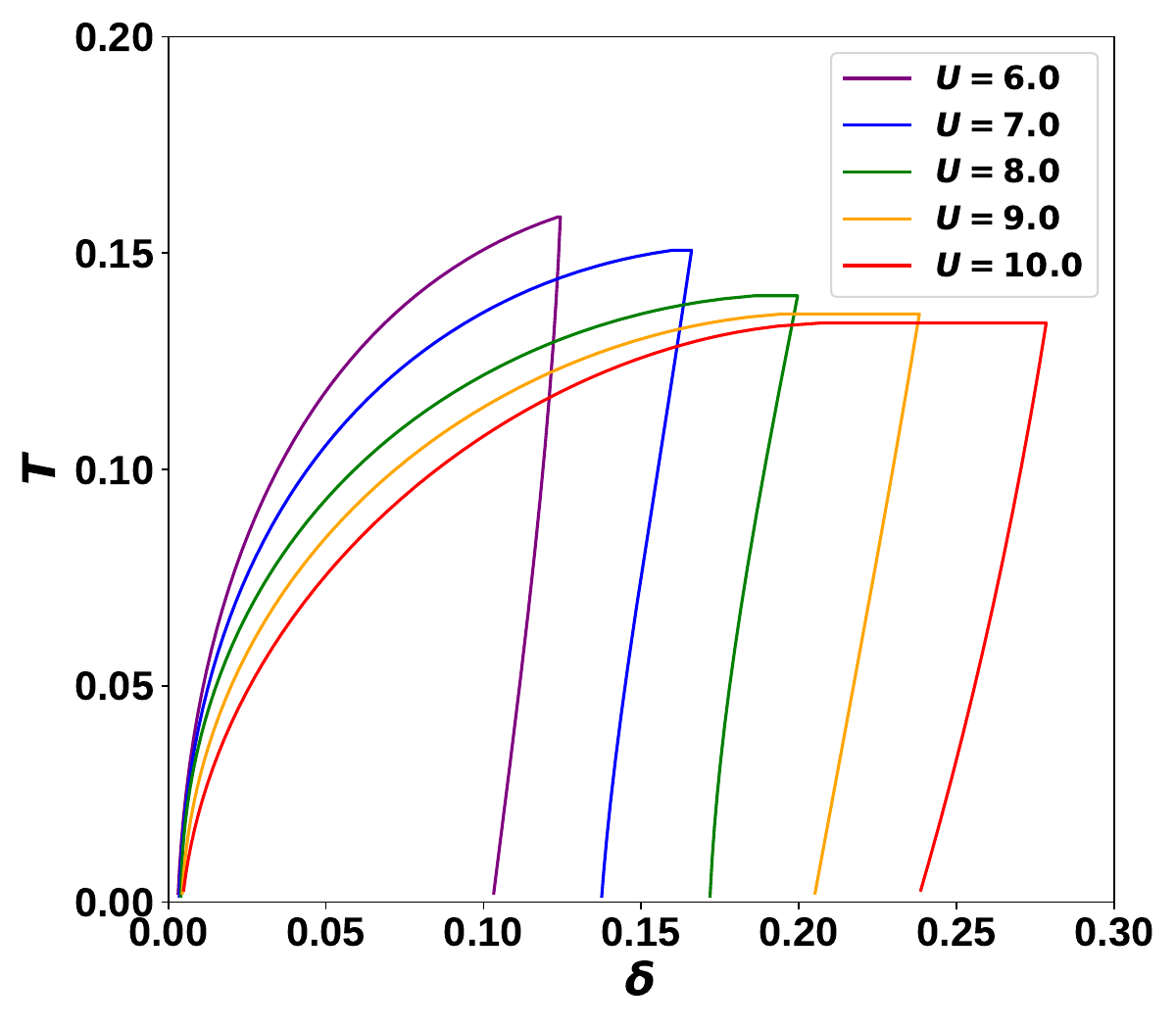}
        \put(18,70){\color{red}\small\bfseries (c)}
    \end{overpic}
\end{subfigure}
\hfill
\begin{subfigure}[t]{0.222\textwidth}
    \centering
    \begin{overpic}[width=\linewidth]{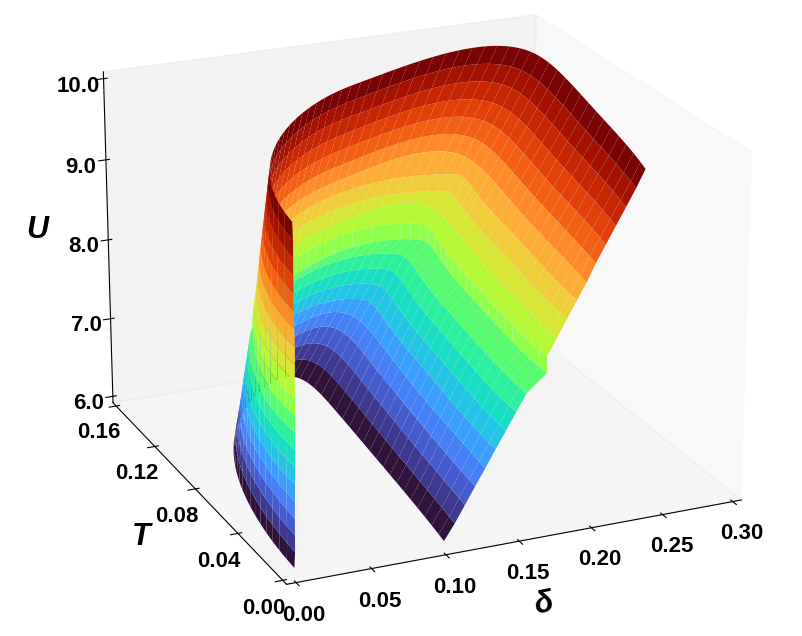}
        \put(14,62){\color{red}\small\bfseries (d)}
    \end{overpic}

\end{subfigure}
\hfill
\begin{subfigure}[t]{0.027\textwidth}
    \centering
    \includegraphics[width=\linewidth,height=3.5cm]{figs/AFM_and_SC_Colorbar.pdf}
\end{subfigure}

\caption{\justifying Interaction-driven evolution of the ordered phases in the temperature-doping plane. (a) AFM phase boundary, (b) SC phase boundary, and (c) PDW phase boundary for several values of \(U\). (d) Three-dimensional visualization of the PDW coexistence region in \(\delta, T, U\) space. The color coding in all panels corresponds to the value of \(U\).}

\label{}

\end{figure}

Fig. 2(c) shows the evolution of the PDW phase boundary for several values of the interaction strength \(U\). As \(U\) increases, the AFM phase expands while the SC dome gradually shrinks. Since the PDW order is generated dynamically through the coexistence and coupling of these two primary orders, its phase boundary reflects the balance between them. Consequently, the PDW region shifts systematically toward higher doping with increasing \(U\), while its overall extent is controlled by the competition between the expanding AFM phase and the diminishing SC phase. Fig. 2(d) presents a three-dimensional visualization of the PDW coexistence region in the (\(\delta,T,U\)) parameter space. The color scale denotes the value of the interaction strength \(U\). This representation provides a comprehensive picture of the evolution of the intertwined AFM+SC+PDW phase as \(U\) is varied. One clearly observes that the coexistence region continuously shifts toward larger doping values with increasing \(U\), while its maximum transition temperature is gradually reduced. The smooth evolution of the surface highlights the intimate connection between the PDW order and the underlying AFM-SC coexistence, confirming that the PDW state is dynamically generated by the interplay of these competing orders rather than arising from an independent instability.  \\

\subsection{Charge-Sector Dynamics and Temperature-Doping Evolution of Order Parameters} 

\begin{figure}[htbp]
\centering

\begin{subfigure}[t]{0.235\textwidth}
    \centering
    \begin{overpic}[width=\linewidth]{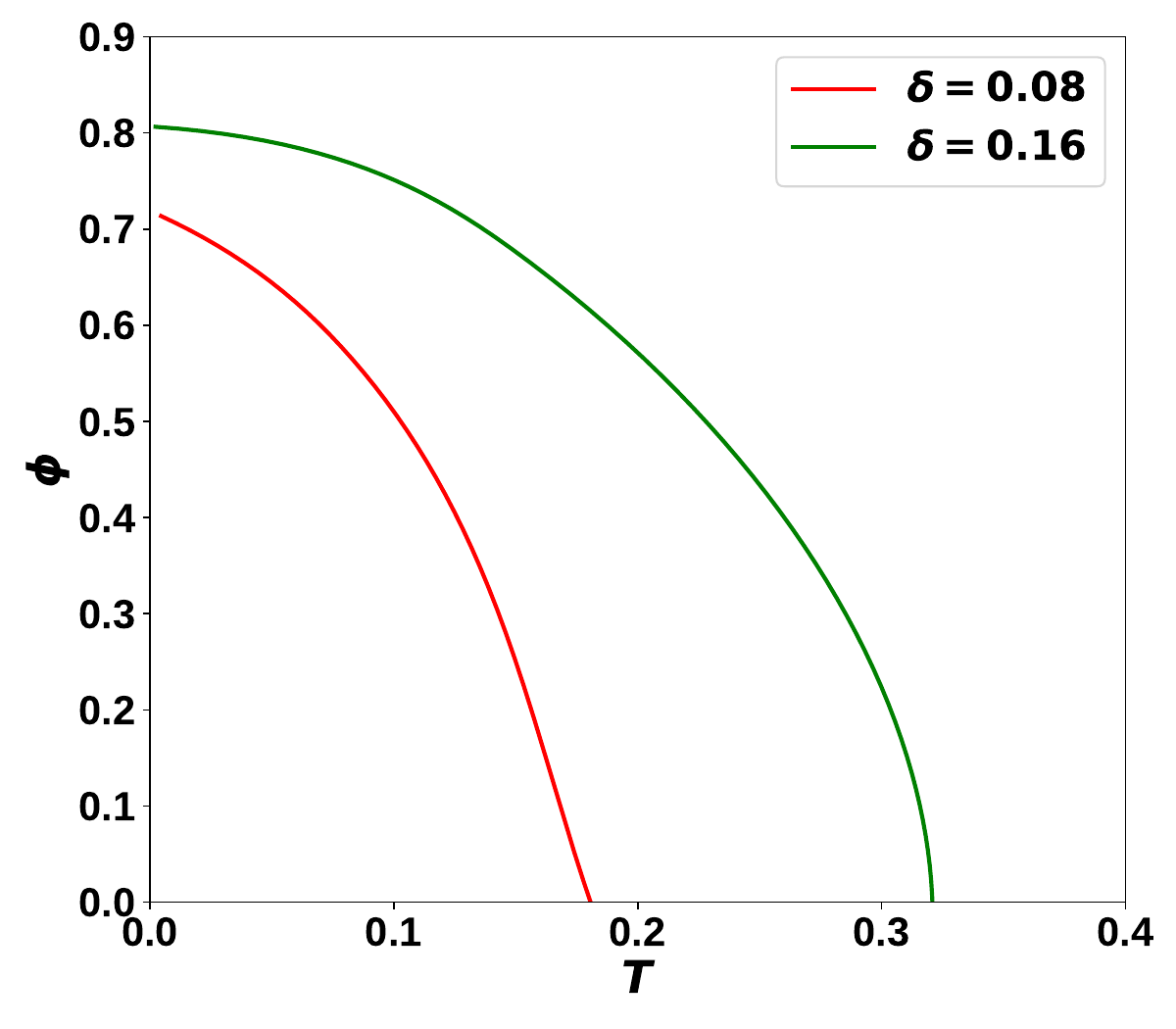}
        \put(63.5,46){
            \includegraphics[width=0.27\linewidth]
            {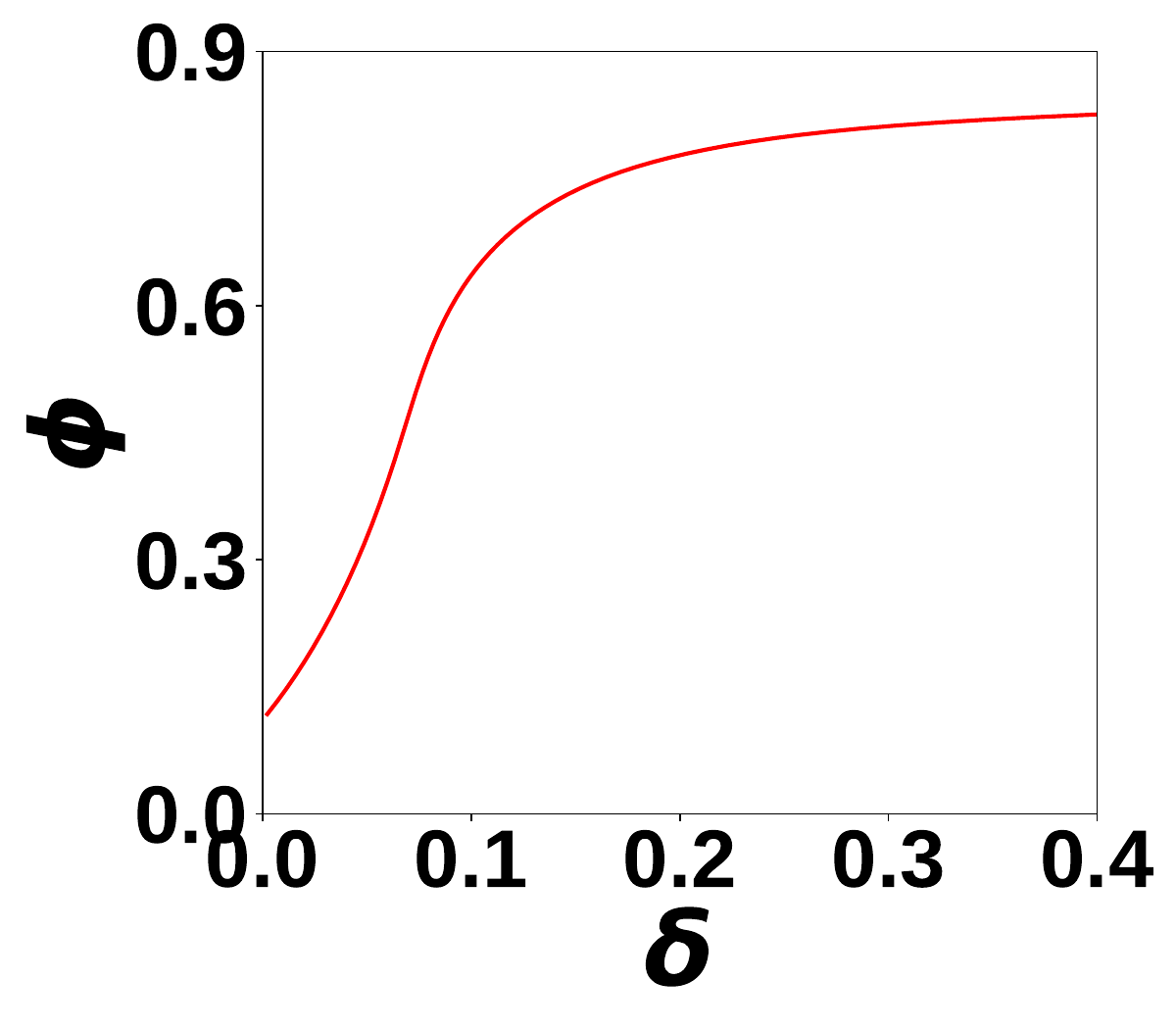}
        }
        \put(50,75){\color{red}\small\bfseries (a)}

    \end{overpic}
    \label{fig:phi}
\end{subfigure}
\hfill
\begin{subfigure}[t]{0.24\textwidth}
    \centering
    \begin{overpic}[width=\linewidth]{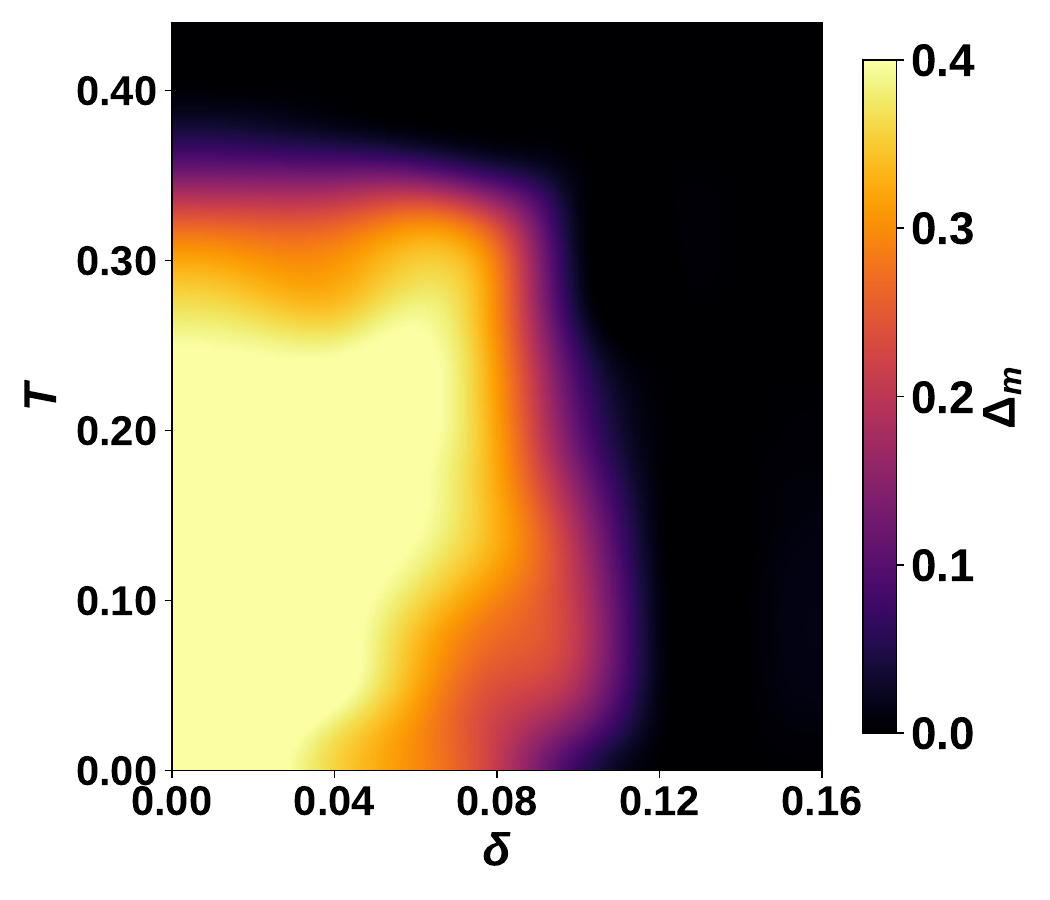}
        \put(64,72){\color{red}\small\bfseries (b)}
    \end{overpic}
    
    \label{fig:afm}
\end{subfigure}

\caption{\justifying (a) Temperature dependence of \(\phi\) for  \(\delta=0.08\) and  \(\delta=0.16\). Inset: Doping dependence of \(\phi\) at fixed temperature \(T=0.08\). (b) Intensity plot of the \(\Delta_m\), in the \(T-\delta\) phase space. The color scale represents the magnitude of \(\Delta_m\).}

\label{fig:phi_afm}

\end{figure}
 
Fig. 3(a) shows the temperature dependence of the rotor condensate \(\phi\) for different values of the doping \(\delta\). The quantity serves as a measure of quantum phase coherence in the rotor sector. As the temperature increases, thermal fluctuations enhance the disorder of the rotor phase \(\theta\), thereby suppressing phase coherence. Consequently, \(\phi\) decreases monotonically with increasing \(T\) and eventually vanishes. For a given temperature, \(\phi\) is larger at higher doping, reflecting the enhanced rotor kinetic energy and the greater stability of charge coherence away from half-filling. \\

In the inset of Fig. 3(a), we show the doping dependence of the rotor condensate \(\phi\) at a fixed temperature \(T=0.08\). In the low-doping regime, \(\phi\) remains strongly suppressed, indicating reduced charge coherence. As the doping \(\delta\) increases, \(\phi\) rises smoothly and gradually approaches saturation value in the high-doping limit. Within the slave–rotor formalism, \(\phi = \langle e^{i\theta}\rangle\) is the order parameter for charge coherence. As doping increases, it introduces the mobile carriers and enhances the quantum charge fluctuations, which promote phase coherence in the rotor sector and lead to a monotonic growth of \(\phi\) with \(\delta\). At higher doping levels, \(\phi\) approaches saturation, indicating that the system has entered into a fully coherent metallic regime where further doping does not significantly enhance charge coherence. In the conventional slave–rotor framework, the evolution of \(\phi\) is typically analyzed as a function of \(U\) at zero temperature and half filling, where it serves as an indicator of the coherence–incoherence transition across the Mott regime \cite{PhysRevB.70.035114, PhysRevB.76.195101, Acharya2016}. \\

 Figure 3(b) presents the intensity map of the AFM order parameter \(\Delta_m\) in the \(T - \delta\) plane. The color scale represents the magnitude of \(\Delta_m\), with brighter colors corresponding to stronger AFM order. \(\Delta_m\) decreases monotonically with increasing doping for a fixed temperature. The introduction of doped holes disrupts the local spin alignment and reduces the effective exchange interactions, thereby weakening the long-range magnetic order. Likewise, for a fixed doping concentration, \(\Delta_m\) decreases with increasing temperature due to the enhancement of thermal fluctuations, which progressively destroy the magnetic correlations. The combined effect of doping-induced charge fluctuations and thermal disorder leads to the gradual suppression of the AFM phase, culminating in a well-defined phase boundary in the \(T - \delta\) plane.

\begin{figure}[htbp]
\centering


\begin{subfigure}[t]{0.22\textwidth}
    \centering
    \begin{overpic}[width=\linewidth]{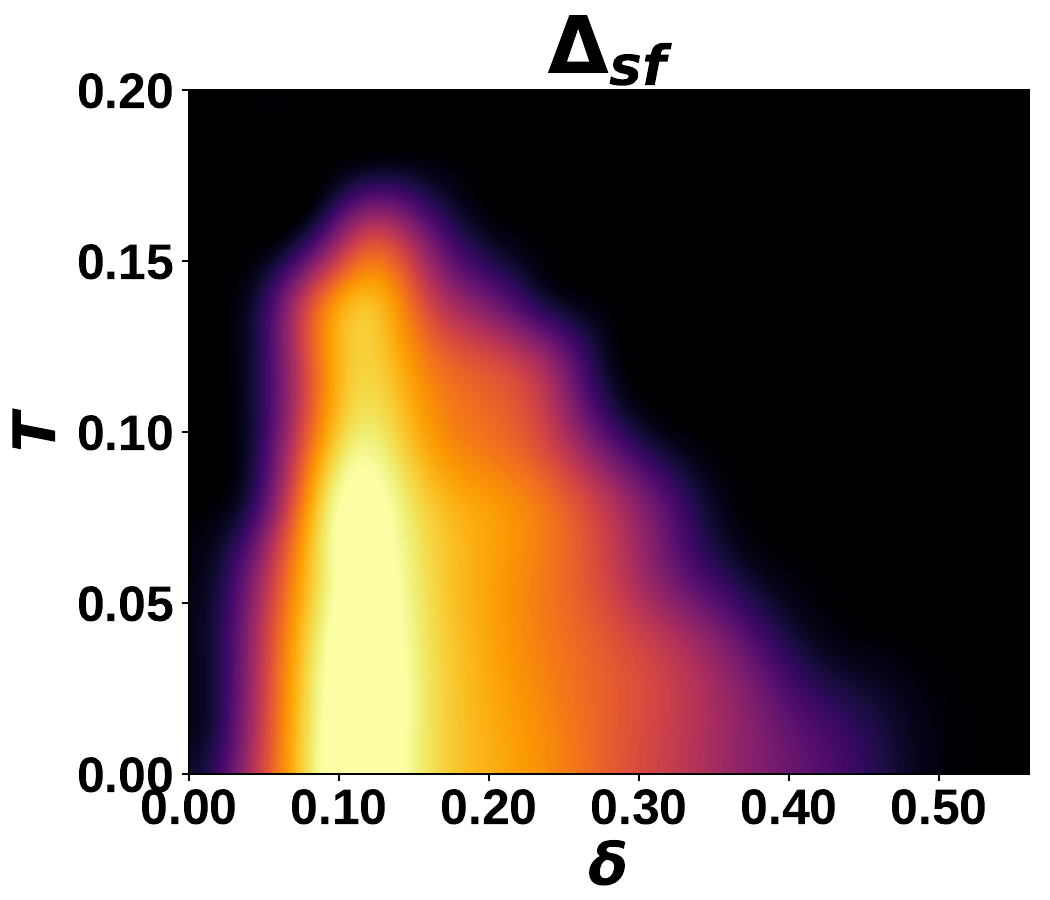}
        \put(80,60){\color{red}\small\bfseries (a)}
    \end{overpic}
\end{subfigure}
\hfill
\begin{subfigure}[t]{0.22\textwidth}
    \centering
    \begin{overpic}[width=\linewidth]{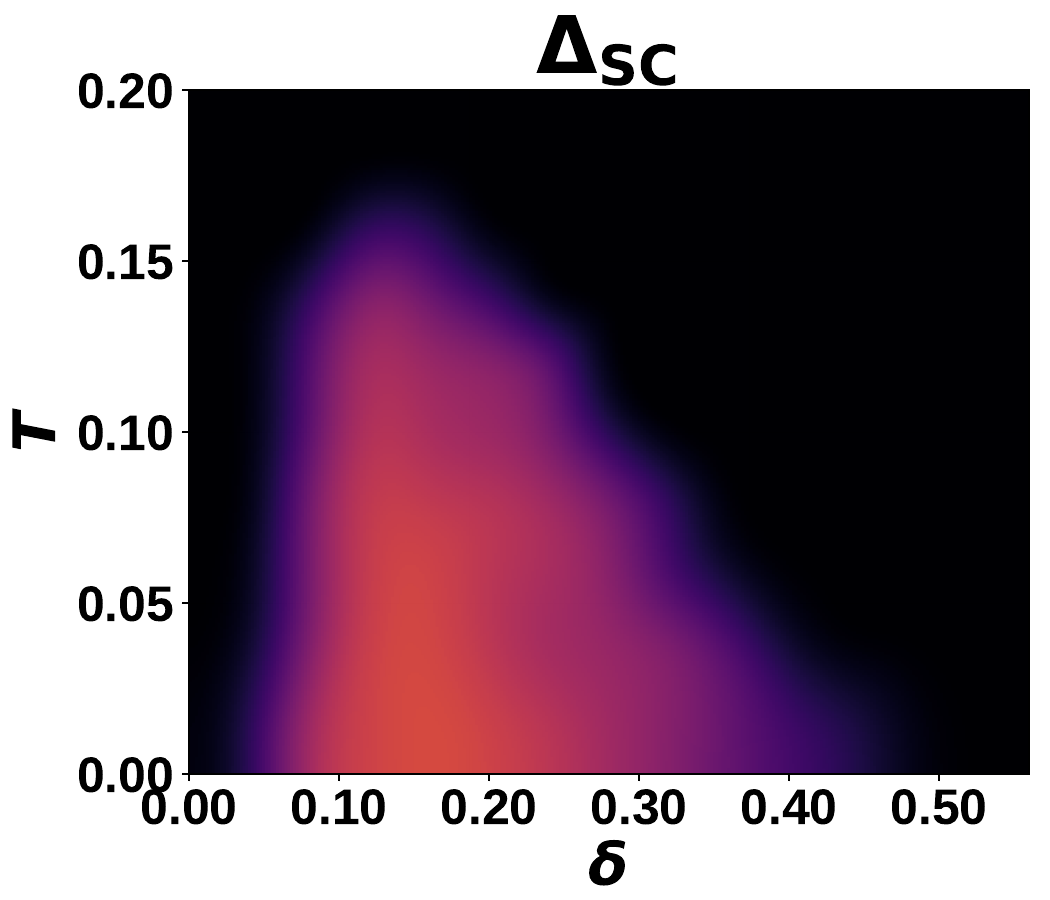}
        \put(80,60){\color{red}\small\bfseries (b)}
    \end{overpic}
\end{subfigure}
\hfill
\begin{subfigure}[t]{0.03\textwidth}
    \centering
    \includegraphics[width=\linewidth,height=3.2cm]{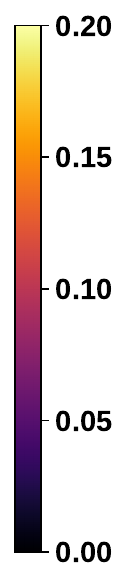}
\end{subfigure}

\vspace{0.05cm}


\begin{subfigure}[t]{0.22\textwidth}
    \centering
    \begin{overpic}[width=\linewidth]{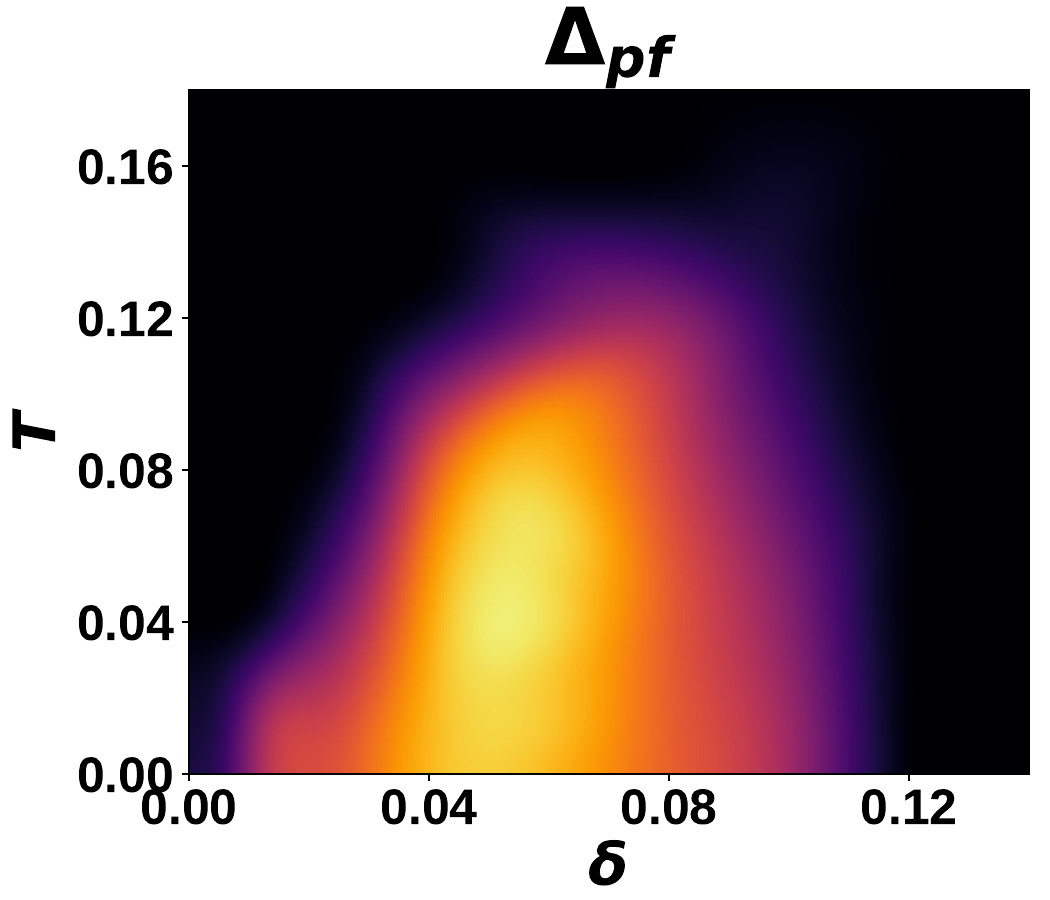}
        \put(80,60){\color{red}\small\bfseries (c)}
    \end{overpic}
    
\end{subfigure}
\hfill
\begin{subfigure}[t]{0.22\textwidth}
    \centering
    \begin{overpic}[width=\linewidth]{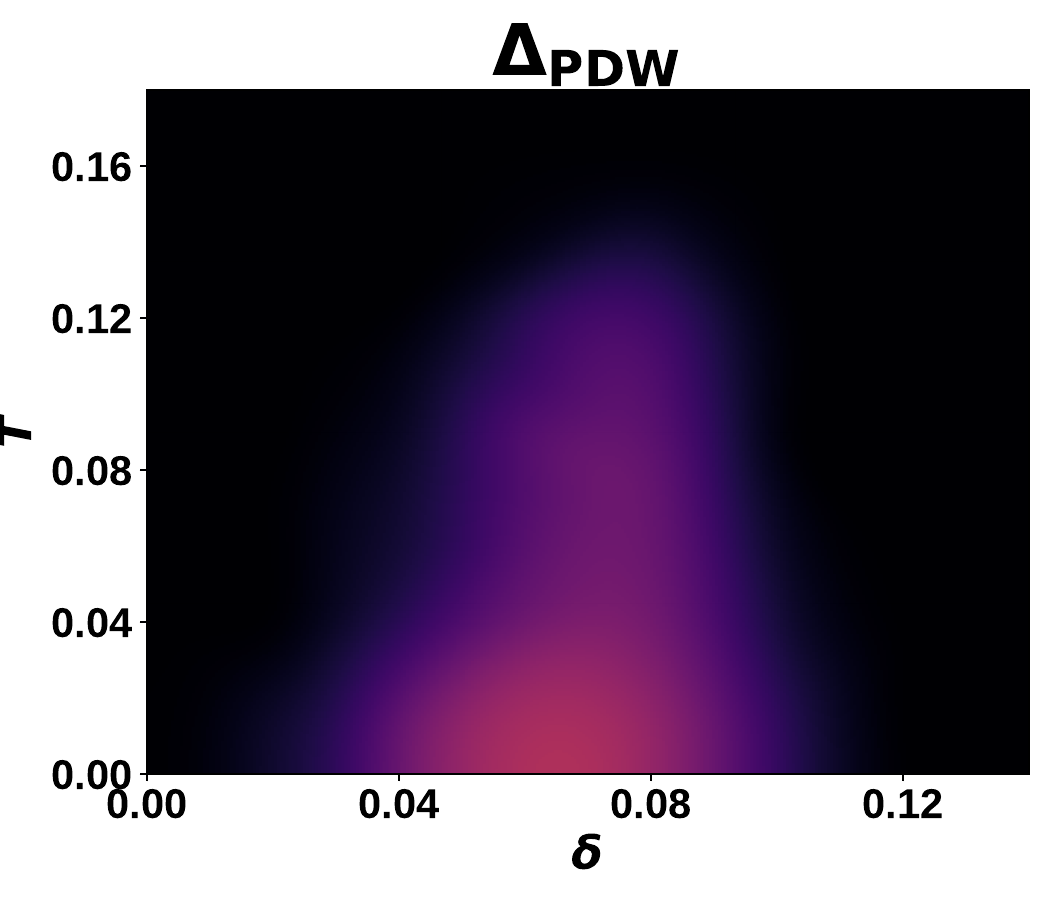}
        \put(80,60){\color{red}\small\bfseries (d)}
    \end{overpic}
\end{subfigure}
\hfill
\begin{subfigure}[t]{0.03\textwidth}
    \centering
    \includegraphics[width=\linewidth,height=3.2cm]{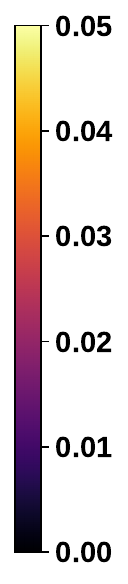}
\end{subfigure}


\caption{\justifying
Intensity plots of the SC and PDW order parameters in the \(T-\delta\) plane. (a) Spinon SC gap parameter \(\Delta_{sf}\). (b) Total SC order parameter \(\Delta_{\mathrm{SC}}=\Delta_{sf}\phi^2\). (c) Spinon PDW gap parameter \(\Delta_{pf}\). (d) Total PDW order parameter \(\Delta_{\mathrm{PDW}}=\Delta_{pf}\phi^2\). The color scales, shown alongside the respective panels, represent the magnitudes of the corresponding order parameters. 
}

\label{fig:SC_PDW_maps}

\end{figure}

Figure 4(a) displays the intensity map of the spinon SC gap parameter \(\Delta_{sf}\) in the \(T-\delta\) plane. At low temperatures, \(\Delta_{sf}\) exhibits a non-monotonic dependence on doping. It is suppressed in the under doped regime due to strong AFM correlations and reduced charge mobility, reaches a maximum at intermediate doping where AFM order is weakened and pairing correlations are strongest, and decreases again in the over doped regime as correlation effects diminish. As the temperature increases, the magnitude of \(\Delta_{sf}\) decreases throughout the phase diagram and the region supporting finite spinon pairing gradually shrinks, reflecting the thermal suppression of the pairing instability.\\

In Fig. 4(b) we show the superconducting order parameter, \(\Delta_{\mathrm{SC}} = \Delta_{sf}\phi^{2}\). We find that \(\Delta_{\mathrm{SC}}\) increases from zero at low doping, reaches a maximum at an intermediate doping level, and decreases gradually in the over doped regime. Importantly, the overall trend closely follows the conventional mean-field behavior of the  superconducting order parameter \cite{PhysRevB.62.9083}. This consistency reflects the fact that \(\phi\) enhances the overall gap scale but does not modify the intrinsic doping dependence arising from the spinon pairing sector, thereby preserving the characteristic mean-field-like evolution of the superconducting gap. The SC region contracts continuously with increasing temperature because thermal fluctuations suppress both the spinon pairing amplitude and the rotor condensate, eventually driving the physical SC order parameter to zero. \\

In Fig. 4(c) and 4(d) we show the intensity plots of the spinon PDW gap \(\Delta_{pf}\) and the corresponding physical PDW order parameter \(\Delta_{\mathrm{PDW}}=\Delta_{pf}\phi^{2}\), respectively. The spinon PDW gap exhibits similar doping dependence as in the case of SC gap, but its support is restricted to a much narrower doping interval. This reflects the fact that PDW order is dynamically generated only in the regime where both the AFM and the singlet SC correlations coexist. With increasing temperature, the magnitude of \(\Delta_{pf}\) decreases and the PDW-supporting region shrinks rapidly. The physical PDW order parameter inherits the same qualitative behavior but is further reduced by the factor \(\phi^{2}\). The finite \(\Delta_{\mathrm{PDW}}\) region is considerably more localized in the \(T - \delta\) plane and disappears at lower temperatures than the spinon PDW gap, emphasizing that the existence of physical PDW order requires not only charge coherence but also spinon PDW correlations which emerge from the coexistence of AFM and SC orders.

\subsection{Interaction-Driven Shift of the tetra-critical Point Temperature}

Fig. 5 shows the dependence of the tetra-critical point temperature \(T_c\) on the antiferromagnetic coupling \(W_m\) and the superconducting coupling \(W_s\). The AFM and SC phases coexist only within the regime \(W_m > W_s\), which defines the parameter window in which the AFM, SC and AFM+SC+PDW phase structure can occur. For \(W_m < W_s\), SC dominates energetically and completely suppresses AFM, eliminating the possibility of coexistence. \\

\begin{figure}[h]
    \centering
    \includegraphics[width=1.0\linewidth]{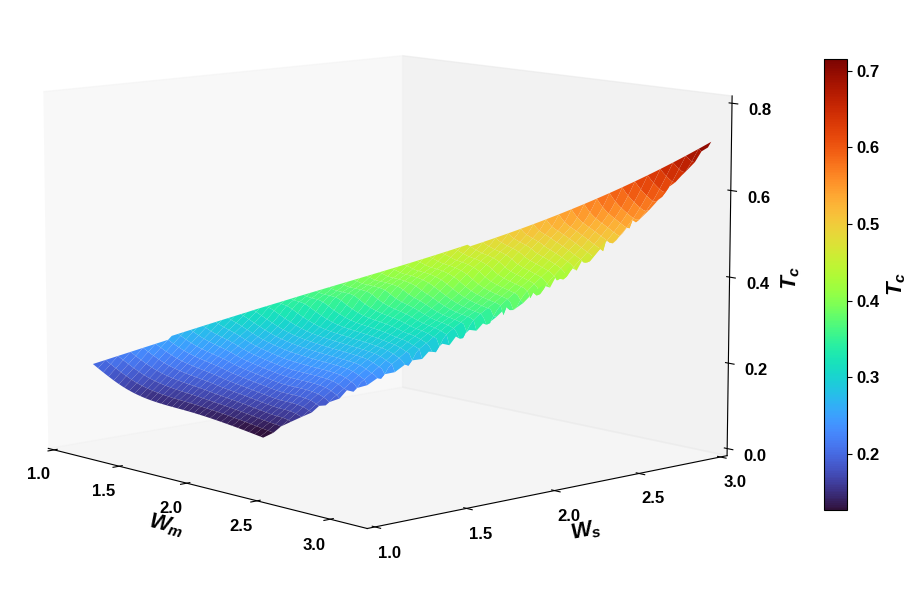}
    \caption{\justifying Color-coded surface plot of the tetracritical point temperature \(T_c\) in the \(W_m - W_s\) parameter space. The color bar indicates the value of \(T_c\), highlighting the interaction-induced shift of the AFM–SC–PDW tetracritical point within the slave–rotor mean-field framework.}
    \label{fig:placeholder}
\end{figure}

For a fixed \(W_s\), \(T_c\) decreases with increasing the \(W_m\). A larger \(W_m\) enhances magnetic order and correspondingly suppresses the superconducting dome. Beyond a certain threshold of \(W_m\), the SC region is pushed entirely inside the AFM sector, eliminating the AFM–SC phase boundary and thereby destroying the tetra-critical point. When \(W_s\) is increased, SC is strengthened, but the AFM coupling must still satisfy \(W_m > W_s\), for the AFM sector to remain competitive, even though the SC dome extends to somewhat higher temperatures. A slight further increase of \(W_m\) strongly favors AFM and pushes the SC region entirely inside the AFM phase. Once this happens, the AFM–SC phase boundary disappears, and the tetra-critical point is lost. Thus, at larger \(W_s\), the allowable range of \(W_m\) that still supports a tetra-critical point becomes very small. Consequently, the strong sensitivity of $T_c$ on \(W_m\) and \(W_s\) sharply constrains the AFM–SC coexistence regime and, in turn, the window for dynamical PDW formation.\\ 

\section{Spectral Features and Density of States} 
Within the slave-rotor mean-field framework, the electron operator is decomposed into charge and spin degrees of freedom. Owing to the mean-field decoupling of the spinon and rotor sectors, the electronic Green's function factorizes approximately into spinon and rotor correlators, 
\begin{equation}
G_c(i,j;\tau) = - \langle T_{\tau}c_{i,\sigma}^{\dagger}(\tau)c_{j,\sigma}(0)\rangle = G_f(i,j;\tau)G_{\theta}(i,j;\tau)
\end{equation} 
\noindent where, \(G_f(i,j;\tau) = -\langle T_{\tau}f_{i,\sigma}^{\dagger}(\tau)f_{j,\sigma}(0)\rangle\) and \(G_{\theta}(i,j;\tau)
= \langle T_{\tau}e^{i\theta_i(\tau)}e^{-i\theta_j(0)}\rangle\) represent the spinon and rotor Green's functions, respectively. \\

After Fourier transformation, the electron Green's function at fermionic Matsubara frequency \(i\omega_n\) is obtained from the convolution of the spinon and rotor Green's functions over the internal momentum \(q\) and bosonic Matsubara frequency \(i\nu_m\).\\

The spinon Green’s function is determined by the
quasiparticle spectrum of the spinon mean-field Hamil-
tonian. It can be expressed in the spectral representation 
\begin{equation}
    G_f(\textbf{k}, i\omega_n) = \sum\limits_{a}\frac{|a,\textbf{k}\rangle \langle a, \textbf{k}|}{i\omega_n - E_{a}(\textbf{k})},
\end{equation}
where, \(E_{a}(\textbf{k})\) are the quasi-particle eigenvalues obtained from the diagonalization of the spinon Hamiltonian and \(a\) labels the quasi-particle bands. \\

The rotor Green's function contains both the condensate contribution and the finite-energy charge fluctuations. The rotor Green's function can be decomposed as 
\begin{equation}
    G_\theta(\textbf{q},i\nu_m) = \beta|\phi|^{2}\delta_{q,0}\delta_{m,0} + G_{\theta}^{\mathrm{inc}}(\textbf{q},i\nu_m), \:\:\: \beta = \frac{1}{kT}
\end{equation}   
Substitution of this decomposition into the electron Green's function naturally separates the electronic spectrum into coherent and incoherent contributions. Consequently, the spectral function can be written as 
\begin{equation}
A(\mathbf{k}, \omega) = A_{\text{coh}}(\textbf{k}, \omega) + A_{\text{inc}}(\mathbf{k}, \omega) 
\end{equation}
The total spectral function satisfies the sum rule \(\int_{-\infty}^{\infty} A(k, \omega) d\omega = 1\). The first term corresponding to the coherent quasi-particle excitations is given by 
\begin{equation}
A_{\text{coh}}(\textbf{k},\omega) = Z  \Bigg[ u_k^{2}\delta(\omega - E_{\mathbf{k}}) + v_k^{2}\delta(\omega + E_{\mathbf{k}})\Bigg], \:\:\: Z = |\phi|^{2}
\end{equation}
which corresponds to the renormalized spinon density of states. \(u_k\) and \(v_k\) are the coherence factors associated with the quasi-particle eigenstates at energies \(\pm E_k\). \(Z\) is an quasiparticle weight.  

\begin{figure*}[t]
    \centering

    \begin{subfigure}[t]{0.32\textwidth}
        \centering
        \begin{overpic}[width=\linewidth]{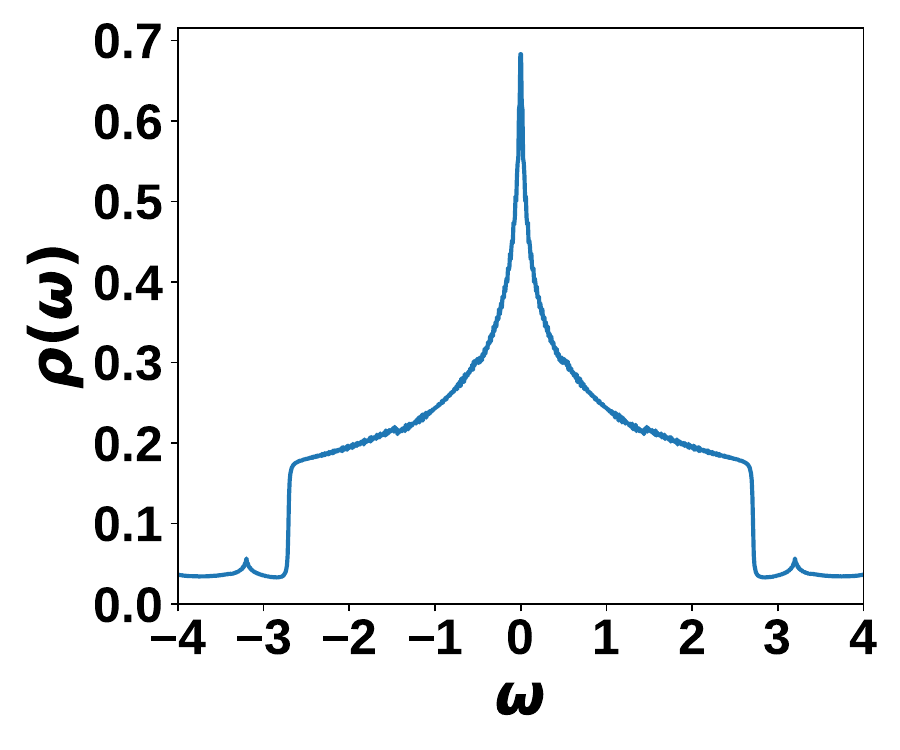}
            \put(83,70){\color{red}\small\bfseries (a)}
        \end{overpic}
    
    \end{subfigure}
    \hfill
    \begin{subfigure}[t]{0.32\textwidth}
        \centering
        \begin{overpic}[width=\linewidth]{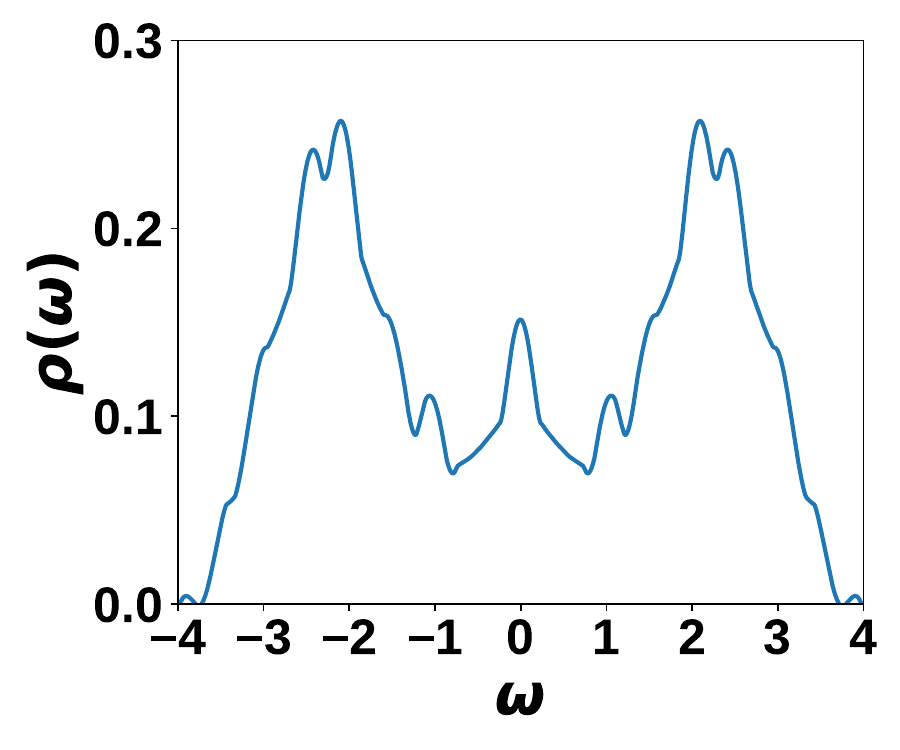}
            \put(83,70){\color{red}\small\bfseries (b)}
        \end{overpic}
    \end{subfigure}
    \hfill
    \begin{subfigure}[t]{0.32\textwidth}
        \centering
        \begin{overpic}[width=\linewidth]{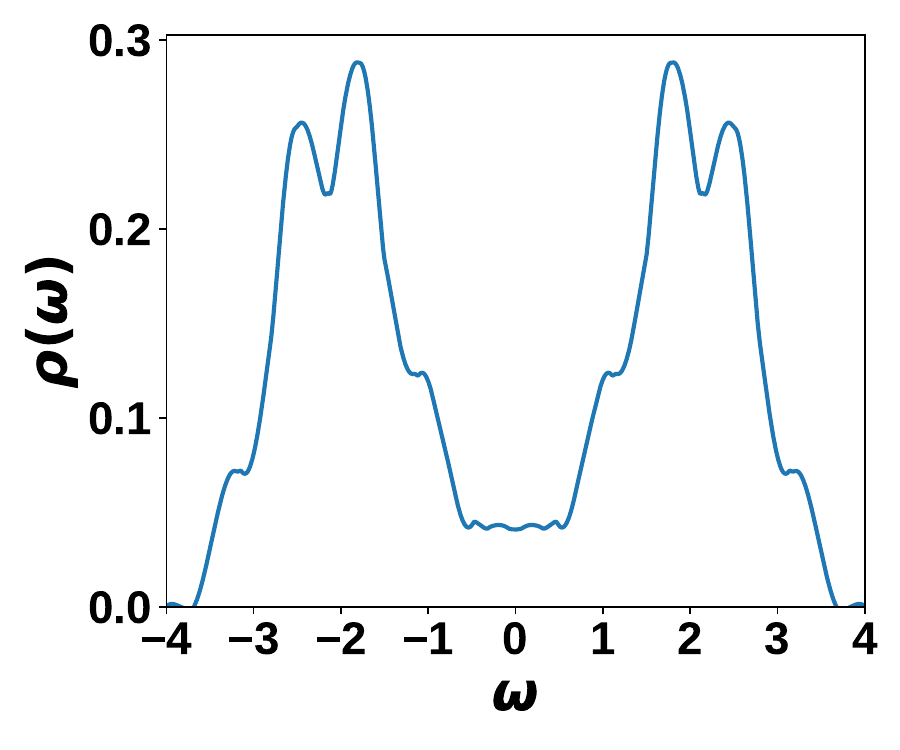}
            \put(83,70){\color{red}\small\bfseries (c)}
        \end{overpic}
    \end{subfigure}


    \caption{\justifying Total density of states (DOS), \(\rho(\omega)\), at fixed doping \(\delta=0.20\) for three representative temperatures: (a) \(T=0.20\) in the coherent metallic phase, (b) \(T=0.36\) in the coherent metallic phase, and (c) \(T=0.40\) corresponding to the crossover from the coherent metal to the incoherent-metal regime. the DOS are plotted relative to the chemical potential.}

    \label{fig:3x3_panel}

\end{figure*}

The incoherent part of the spectral function arises from finite-energy excitations of the rotor sector and is given by
\begin{equation}
\begin{aligned}
A_{\text{inc}}(\mathbf{k}, \omega) = &\sum_{m,n, m \neq n} \Bigg[u_k^{2}
\frac{e^{-\beta w_m}}{Z_{\text{par}}}
\left|\langle \psi_m | e^{-i\theta} | \psi_n \rangle \right|^{2}\\
&\delta\big(\omega - (E_{\mathbf{k}} + w_{mn})\big) +
v_k^{2}\frac{e^{-\beta w_n}}{Z_{\text{par}}} \\
&\left|\langle \psi_n | e^{-i\theta} | \psi_m \rangle \right|^{2} 
\delta\big(\omega - (-E_{\mathbf{k}} + w_{mn})\big) 
\Bigg],
\end{aligned}
\end{equation} 
where \(|\psi_m\rangle\) and \(|\psi_n \rangle\) are the eigenstates of the rotor Hamiltonian and $w_{mn}=w_{m} - w_{n}$ is the excitation energy involving rotor energy states \(w_m\) and \(w_n\) and \(Z_{par}\) is the rotor partition function. The incoherent spectral function originates from transitions between different rotor eigenstates and captures finite-energy charge fluctuations beyond the coherent quasiparticle contribution. The thermally averaged probabilities for charge-transfer processes in the rotor sector are \(|\langle \psi_m|e^{-i \theta}|\psi_n\rangle|^{2}\). These processes shift the spinon dispersion by the rotor excitation energies \(w_m - w_n\), giving rise to incoherent features at finite frequencies. \\

Finally, the total density of states is given by
\begin{equation}
\rho(\omega) = \rho_{\text{coh}}(\omega) + \rho_{\text{inc}}(\omega).
\end{equation}
This formulation allows us to analyze the redistribution of spectral weight across different regions of the phase diagram, distinguishing between coherent metallic and incoherent-metal regimes. 

The above formulation provides a unified framework for analyzing the spectral properties throughout the phase diagram. In particular, the evolution of coherent and incoherent spectral weight enables a direct distinction between ordered phases, coherent metallic states, and incoherent-metal regimes, thereby revealing the role of charge fluctuations in the redistribution of low-energy electronic spectral weight.

\subsection{Coherent and Incoherent Metal}

In Fig. 6, we present the total density of states, including both coherent and incoherent contributions, at fixed doping for different temperatures. At low temperature (\(\phi = 0.80, T=0.20\)), the coherent contribution dominates the spectral function, yielding a large quasi-particle weight \(Z = \phi^{2}\) and a pronounced peak at the Fermi level, (\(\omega = 0\)). Upon increasing the temperature (\(\phi =0.40, T=0.36\)), the quasi-particle weight is substantially reduced, leading to a suppression of the coherent peak and the emergence of incoherent sidebands at finite energies. This behavior reflects a redistribution of spectral weight from the coherent quasi-particle sector to incoherent excitations. In the intermediate-temperature regime, the system remains metallic, characterized by the coexistence of residual coherent quasi-particles and significant incoherent contributions. At higher temperature \((\phi=0, T=0.40)\), the coherent contribution vanishes entirely, and the spectrum is dominated by two well-separated incoherent peaks. Notably, although the quasiparticle peak at \(\omega = 0\) disappears, the density of states at the Fermi level remains finite, indicating the absence of a gap. This behavior is characteristic of an incoherent-metal state, where well-defined quasiparticles are absent despite finite low-energy spectral weight.  \\

\begin{figure}[htbp]
\centering


\begin{subfigure}[t]{0.222\textwidth}
    \centering
    \begin{overpic}[width=\linewidth]{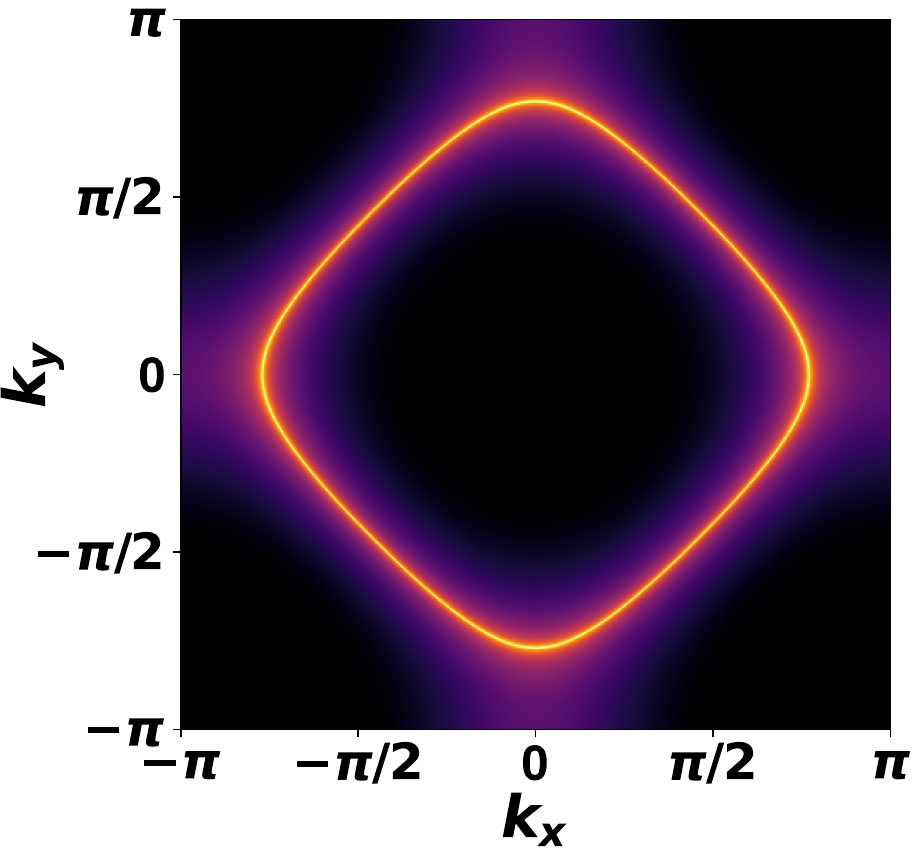}
        \put(80,80){\color{red}\small\bfseries (a)}
    \end{overpic}
    
\end{subfigure}
\hfill
\begin{subfigure}[t]{0.222\textwidth}
    \centering
    \begin{overpic}[width=\linewidth]{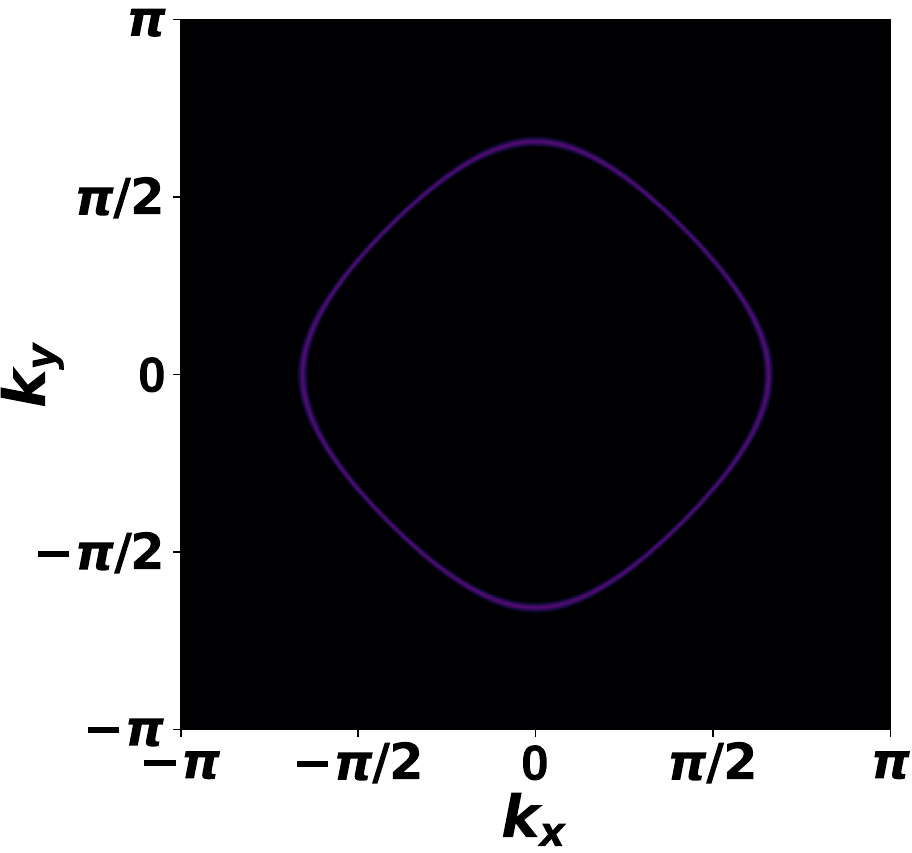}
        \put(80,80){\color{red}\small\bfseries (b)}
    \end{overpic}
\end{subfigure}
\hfill
\begin{subfigure}[t]{0.027\textwidth}
    \centering
    \includegraphics[width=\linewidth,height=3.6cm]{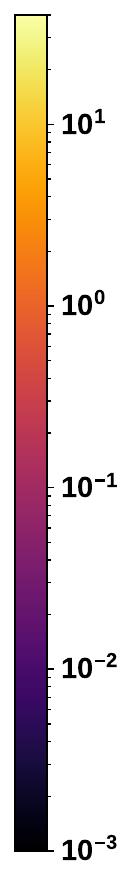}
\end{subfigure}

\caption{\justifying Intensity maps of the zero-energy spectral function \(A(\mathbf{k},\omega=0)\) in the two-dimensional Brillouin zone at fixed doping \(\delta=0.20\). (a) Coherent metallic phase at \(T=0.20\), with \(\phi=0.80\). (b) Near the incoherent-metal regime at \(T=0.39\), with \(\phi=0.01\). The color scale (logarithmic) represents the magnitude of \(A(\mathbf{k},\omega=0)\).}

\label{}

\end{figure}

Fig. 7 shows the momentum-resolved zero-energy coherent spectral function \(A_{\mathrm{coh}}(\mathbf{k},\omega=0)\) for two representative temperatures at fixed doping \((\delta=0.20)\). The zero-energy coherent spectral function remains sharply localized on the Fermi surface. Since the coherent spectral weight is proportional to the quasi-particle residue \((Z=\phi^{2})\), the bright contour directly represents the coherent Fermi surface. At low temperature \(T=0.20\), \(\phi=0.80\), a well-defined Fermi-surface contour with strong spectral intensity is observed, characteristic of coherent quasi-particle excitations. As the \(T\) increases to \(T=0.39\), the rotor condensate is strongly suppressed to   

\begin{figure*}[htbp]
    \centering

    \begin{subfigure}[t]{0.308\textwidth}
        \centering
        \begin{overpic}[width=\linewidth]{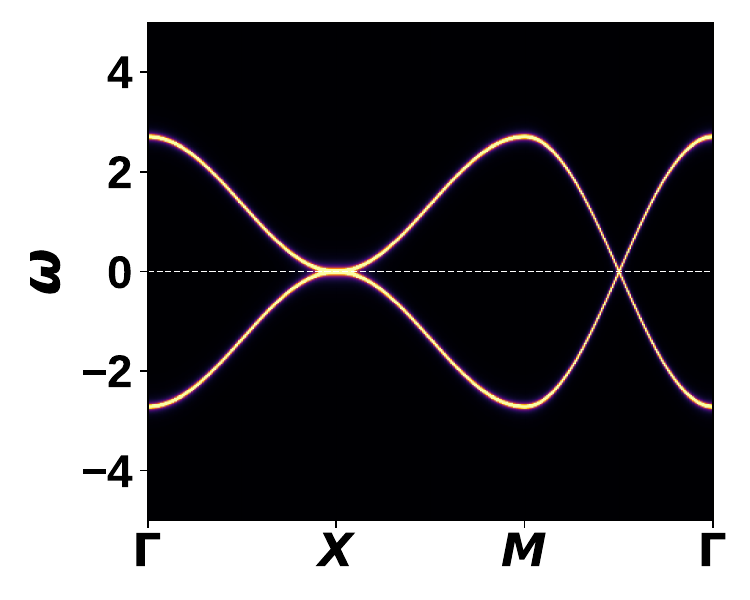}
            \put(85,70){\color{red}\small\bfseries (a)}
        \end{overpic}

    \end{subfigure}
    \hfill
    \begin{subfigure}[t]{0.308\textwidth}
        \centering
        \begin{overpic}[width=\linewidth]{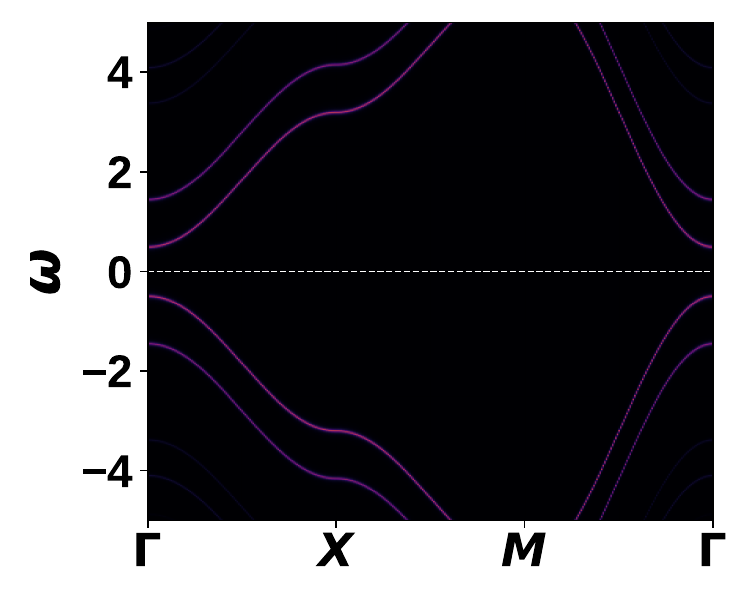}
            \put(85,70){\color{red}\small\bfseries (b)}
        \end{overpic}
        
    \end{subfigure}
    \hfill
    \begin{subfigure}[t]{0.308\textwidth}
        \centering
        \begin{overpic}[width=\linewidth]{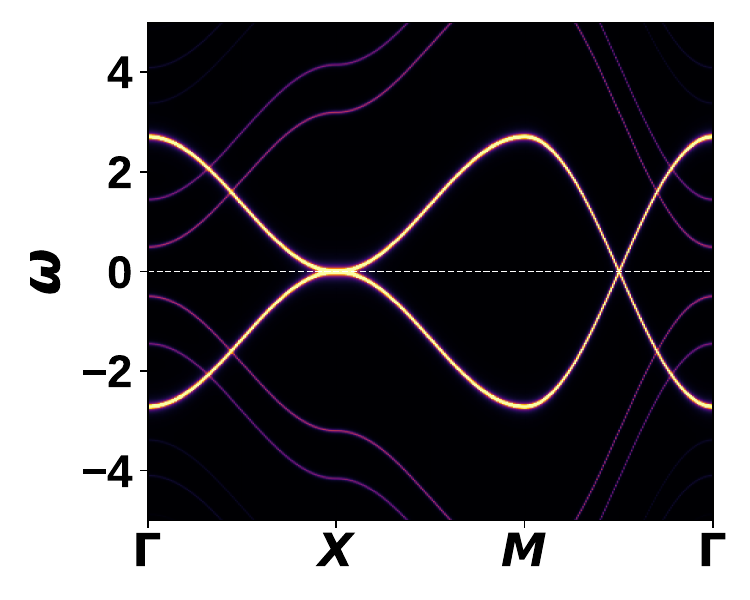}
            \put(85,70){\color{red}\small\bfseries (c)}
        \end{overpic}
        
    \end{subfigure}
    \hfill
    \begin{subfigure}[t]{0.03\textwidth}
        \centering
        \includegraphics[width=\linewidth,height=4.2cm]{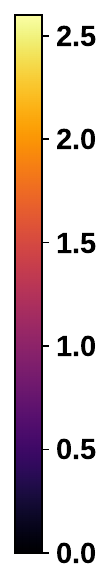}
    \end{subfigure}


    \begin{subfigure}[t]{0.308\textwidth}
        \centering
        \begin{overpic}[width=\linewidth]{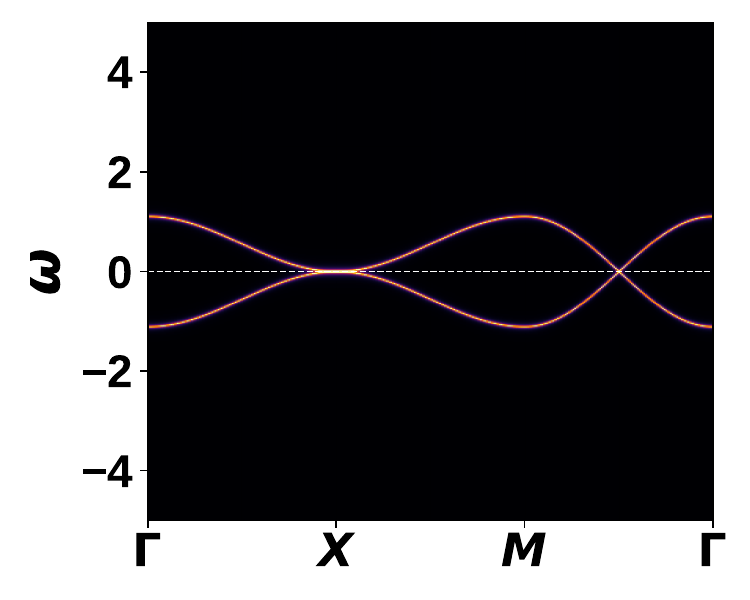}
            \put(85,70){\color{red}\small\bfseries (d)}
        \end{overpic}
    \end{subfigure}
    \hfill
    \begin{subfigure}[t]{0.308\textwidth}
        \centering
        \begin{overpic}[width=\linewidth]{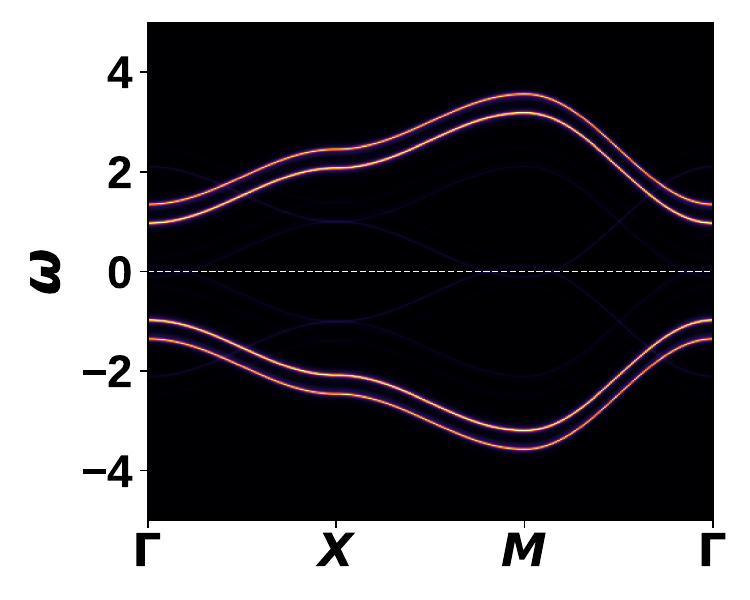}
            \put(85,70){\color{red}\small\bfseries (e)}
        \end{overpic}

    \end{subfigure}
    \hfill
    \begin{subfigure}[t]{0.308\textwidth}
        \centering
        \begin{overpic}[width=\linewidth]{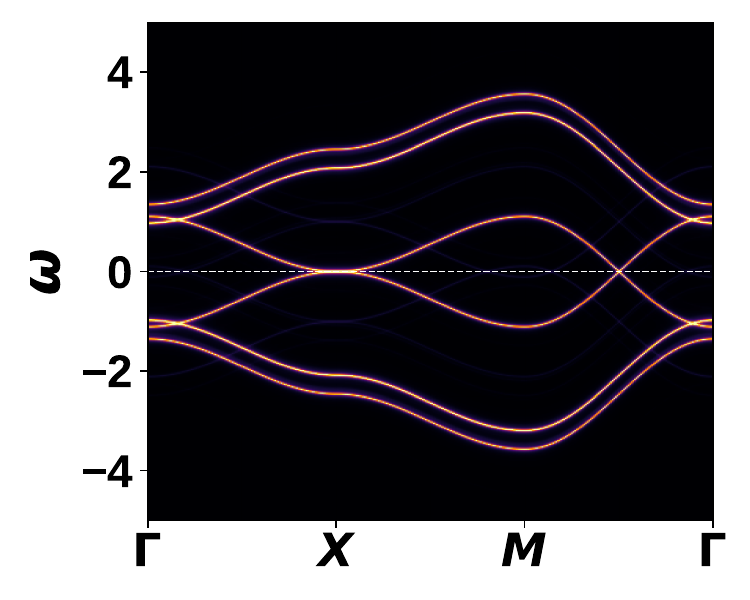}
            \put(85,70){\color{red}\small\bfseries (f)}
        \end{overpic}
 
    \end{subfigure}
    \hfill
    \begin{subfigure}[t]{0.03\textwidth}
        \centering
        \includegraphics[width=\linewidth,height=4.2cm]{figs/Coherent_to_Bad_metal.pdf}
    \end{subfigure}


    \begin{subfigure}[t]{0.308\textwidth}
        \centering
        \begin{overpic}[width=\linewidth]{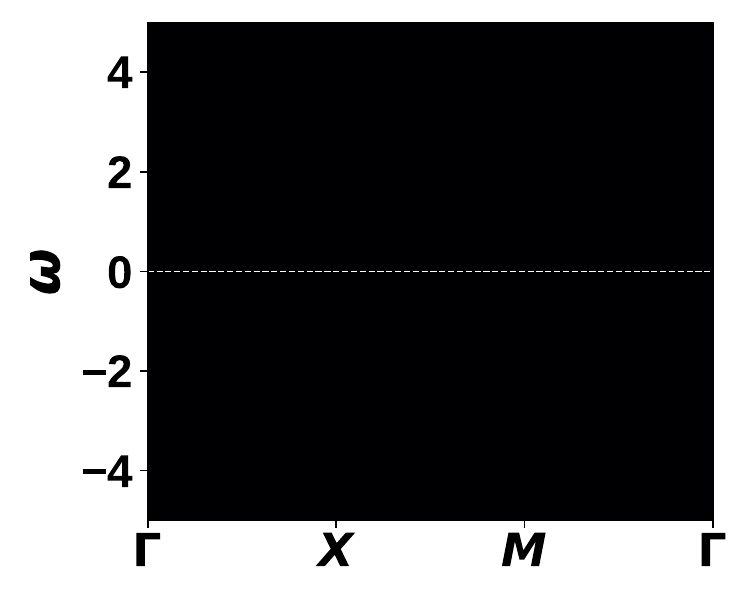}
            \put(85,70){\color{red}\small\bfseries (g)}
        \end{overpic}

    \end{subfigure}
    \hfill
    \begin{subfigure}[t]{0.308\textwidth}
        \centering
        \begin{overpic}[width=\linewidth]{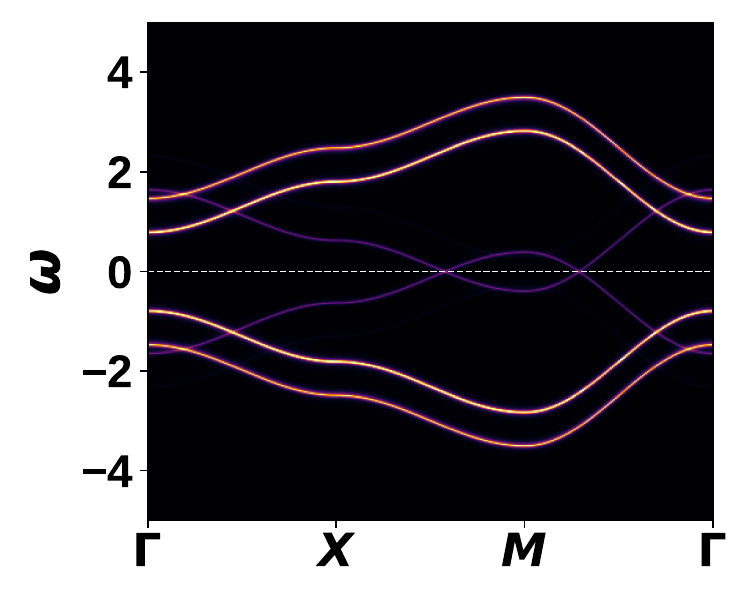}
            \put(85,70){\color{red}\small\bfseries (h)}
        \end{overpic}
    \end{subfigure}
    \hfill
    \begin{subfigure}[t]{0.308\textwidth}
        \centering
        \begin{overpic}[width=\linewidth]{figs/A_k_omega_for_NEW_total_and_incoherent_for_T_=_0.40.pdf}
            \put(85,70){\color{red}\small\bfseries (i)}
        \end{overpic}

    \end{subfigure}
    \hfill
    \begin{subfigure}[t]{0.03\textwidth}
        \centering
        \includegraphics[width=\linewidth,height=4.2cm]{figs/Coherent_to_Bad_metal.pdf}
    \end{subfigure}


    \caption{\justifying Momentum-resolved spectral function \(A(\mathbf{k},\omega)\) along the high-symmetry path for fixed hole doping \(\delta=0.20\). The first, second, and third columns show the coherent, incoherent, and total spectral functions, respectively. The three rows correspond to temperatures (a)--(c) \(T=0.20\) in the coherent metallic phase, (d)--(f) \(T=0.36\) in the coherent metallic phase, and (g)--(i) \(T=0.40\) corresponding to the crossover from the coherent metal to the incoherent-metal regime. The spectra are plotted with respect to the chemical potential.}

    \label{fig:3x3_panel}

\end{figure*}

\noindent
\(\phi=0.01\), leading to a substantial reduction of the coherent spectral weight. Although the underlying Fermi-surface contour remains discernible, its intensity is greatly diminished, indicating that the coherent quasiparticles progressively lose their spectral weight as the system evolves toward the incoherent-metal regime. \\

To further characterize the coherent-to-incoherent crossover, we exmine the evolution of the quasiparticle effective mass obtained from the quasiparticle weight through the relation \(m^{*}/m_0 = 1/Z = 1/ \phi^{2}\). At low temperature \(T=0.20\), the effective mass is only moderately enhanced \(m^{*}/m_0 \approx 1.56\), consistent with the pronounced coherent peak at the Fermi level and well-defined quasiparticles. With increasing temperature \(T=0.36\), the effective mass increases substantially \(m^{*}/m_0 = 6.25\). The pronounced enhancement of the effective mass indicates that the quasiparticles become progressively heavier with increasing temperature, reflecting the gradual loss of coherence and the evolution toward an incoherent metallic state. Near the crossover, the effective mass becomes very large, reflecting the collapse of coherent quasiparticle dynamics and the emergence of an incoherent metallic state. Thus, the strong enhancement of the effective mass provides a clear signature of the coherent-to-incoherent crossover. \\

Fig. 8 shows the momentum-resolved coherent, incoherent, and total spectral functions along the \(\Gamma\)-\(X\)-\(M\)-\(\Gamma\) path for three representative temperatures and at fixed hole doping \(\delta=0.20\). At low temperature \(T=0.20\), the total spectrum is dominated by coherent quasi-particle bands crossing the Fermi level near the \(X\) point, characteristic of a coherent metallic state with well-defined quasiparticle excitations and indicating that the Fermi surface is intersected in this region of momentum space.  With increasing temperature \(T=0.36\), the coherent spectral weight is significantly reduced while the incoherent contribution becomes increasingly prominent, indicating a transfer of spectral weight from coherent quasiparticles to incoherent excitations. At \(T=0.40\), where the rotor condensate vanishes, the coherent bands disappear completely and the total spectral function is governed entirely by the incoherent sector. The incoherent spectrum exhibits a characteristic pairwise band structure on both sides of the Fermi level. The pairwise bands observed on both sides of the Fermi level arise from the splitting of an single incoherent band due to the presence of intersite rotor correlations in the rotor Hamiltonian.

\subsection{AFM + (\(\phi \neq 0\)) and AFM + (\(\phi = 0\))} 

Now, we observe that there are two regions inside the AFM regions - one is AFM + (\(\phi \neq 0 \)) and another one is AFM + (\(\phi = 0\)). In Fig. 9, we show total DOS plot for both regions.
In AFM + (\(\phi \neq 0\)), there are coherent and incoherent part which contributes to the total DOS plot. The coherent part has a gap in the fermi level and two peaks are located at \(\omega = \pm W_m\Delta_m\) separated by the gap. We see the Hubbard-like bands are originated due to incoherent part of the spectral function. In the present formulation, the interaction term is effectively partitioned between the spinon and rotor sectors, with the rotor Hamiltonian governed by an energy scale \(U/2\). Consequently, the dominant rotor excitations occur at energies of order \(U/2\), leading to Hubbard-like bands centered around \(\omega \approx \frac{U}{2}\). \\

\begin{figure}[h!]
\centering


\begin{subfigure}[t]{0.23\textwidth}
    \centering
    \begin{overpic}[width=\linewidth]{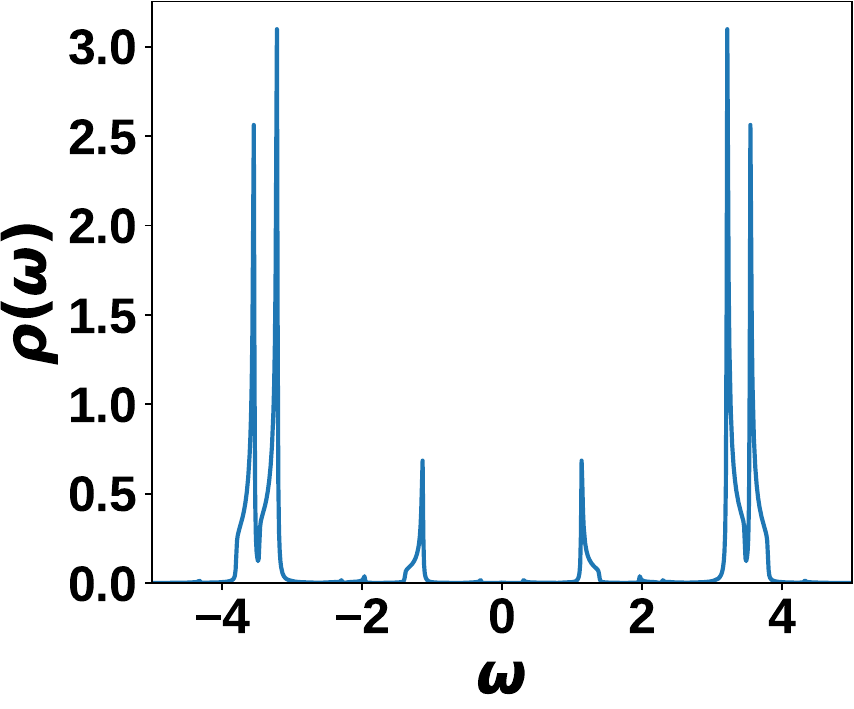}
        \put(88,70){\color{red}\small\bfseries (a)}
    \end{overpic}
    
\end{subfigure}
\hfill
\begin{subfigure}[t]{0.23\textwidth}
    \centering
    \begin{overpic}[width=\linewidth]{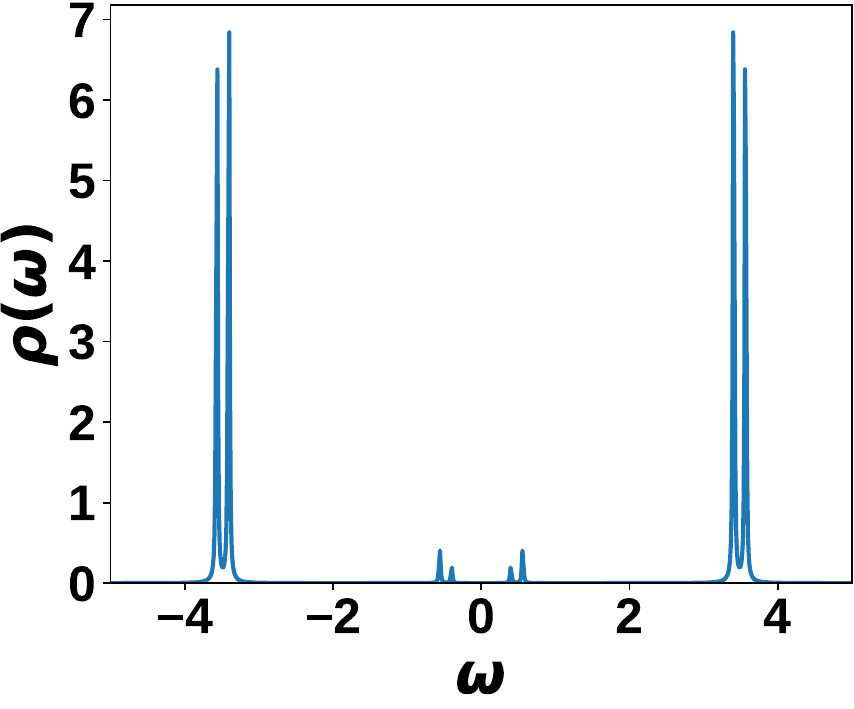}
        \put(89,70){\color{red}\small\bfseries (b)}
    \end{overpic}
    
\end{subfigure}

\caption{\justifying Total density of states for (a) the AFM+\((\phi \neq 0)\) phase at \(T=0.16\) and \(\delta=0.10\), and (b) the AFM+\((\phi=0)\) phase at \(T=0.24\) and \(\delta=0.08\). The chosen temperature and doping values belong to the respective regions identified in the phase diagram.}

\end{figure} 

A prominent feature of the incoherent spectrum is the pairwise splitting of the Hubbard-like bands on both sides of the Fermi level. To elucidate the origin of this structure, we examine the role of the intersite rotor-correlation term \( -2t\chi(e^{-i\theta_1}e^{i\theta_2}+\mathrm{h.c.})\) in the rotor Hamiltonian. This term couples rotor excitations residing on different sites and  consequently hybridizes degenerate local rotor states. The resulting level repulsion lifts the degeneracy of the rotor excitation spectrum, causing a single incoherent band to split into two distinct branches. As a consequence, the Hubbard-like features appear as pairwise peaks in both the momentum-resolved spectral function and the density of states.\\

To gain further insight into the nature of the excitations, we examine the momentum-resolved coherent and incoherent spectral functions along the \(\Gamma\)-\(X\)-\(M\)-\(\Gamma\)  path, as shown in Fig. 10. The coherent spectrum consists of two well-defined quasiparticle branches separated by a finite gap at the Fermi level, reflecting the antiferromagnetic order present in the AFM+\((\phi \neq 0)\) phase. These branches correspond to the low-energy spinon quasiparticles dressed by the finite rotor condensate and are responsible for the sharp coherent peaks observed in the total density of states.

\begin{figure}[htbp]
\centering


\begin{subfigure}[t]{0.222\textwidth}
    \centering
    \begin{overpic}[width=\linewidth]{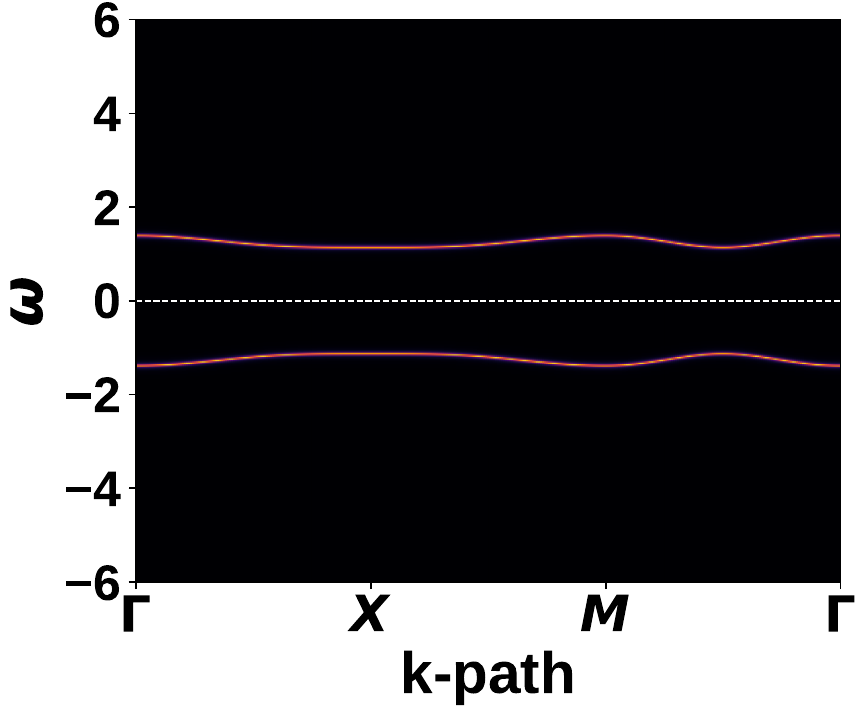}
        \put(85,70){\color{red}\small\bfseries (a)}
    \end{overpic}
    
\end{subfigure}
\hfill
\begin{subfigure}[t]{0.222\textwidth}
    \centering
    \begin{overpic}[width=\linewidth]{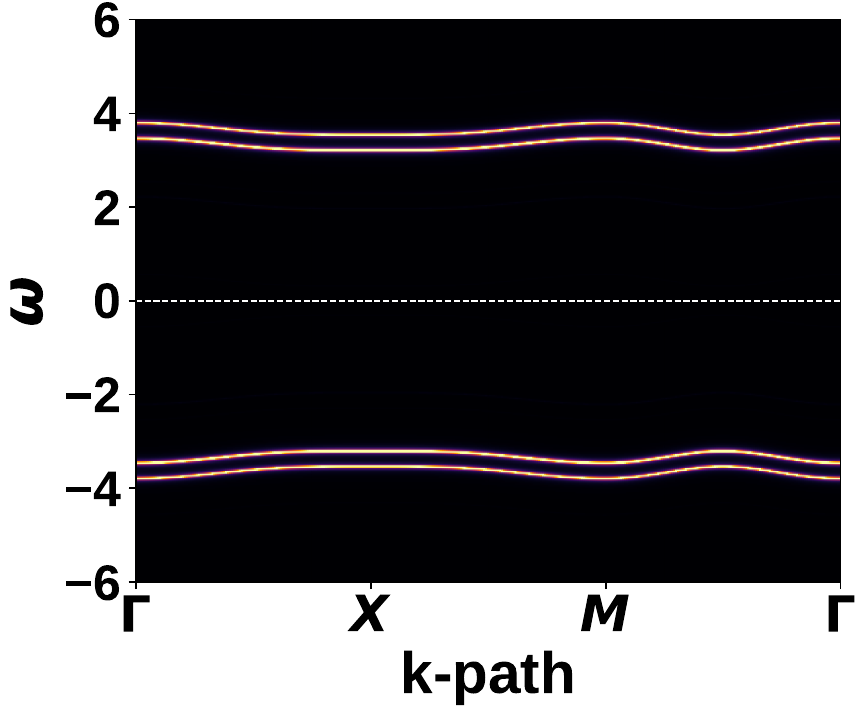}
        \put(85,70){\color{red}\small\bfseries (b)}
    \end{overpic}
    
\end{subfigure}
\hfill
\begin{subfigure}[t]{0.027\textwidth}
    \centering
    \includegraphics[width=\linewidth,height=3.2cm]{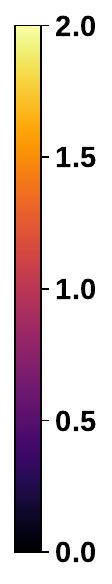}
\end{subfigure}

\caption{\justifying Momentum-resolved (a) coherent and (b) incoherent spectral functions \(A(\textbf{k},\omega)\) along the \(\Gamma\)-\(X\)-\(M\)-\(\Gamma\) path corresponding to the AFM+\((\phi \neq 0)\) phase state shown in Fig. 9(a).}

\end{figure} 

In contrast, the incoherent spectrum is composed of higher-energy excitations arising from the rotor sector. Rather than forming a single dispersive feature, the dominant incoherent bands appear as closely spaced pairs on both sides of the Fermi level. This pairwise structure is a direct consequence of the intersite rotor-correlation term in the rotor Hamiltonian, which couples rotor excitations on neighboring sites and removes the degeneracy of the corresponding excitation energies. The resulting hybridization leads to a splitting of the incoherent branches, producing the characteristic double-band structure visible throughout the momentum path. The same splitting is reflected in the density of states as pairwise Hubbard-like peaks. We further verify this interpretation by suppressing the intersite rotor-correlation term, in which case the split branches collapse into a single incoherent band. These results demonstrate that intersite rotor correlations play a crucial role in determining the structure of the incoherent spectral weight in the AFM phase. 

\subsection{SC and Coexistence}

\begin{figure}[htbp]
\centering


\begin{subfigure}[t]{0.23\textwidth}
    \centering
    \begin{overpic}[width=\linewidth]{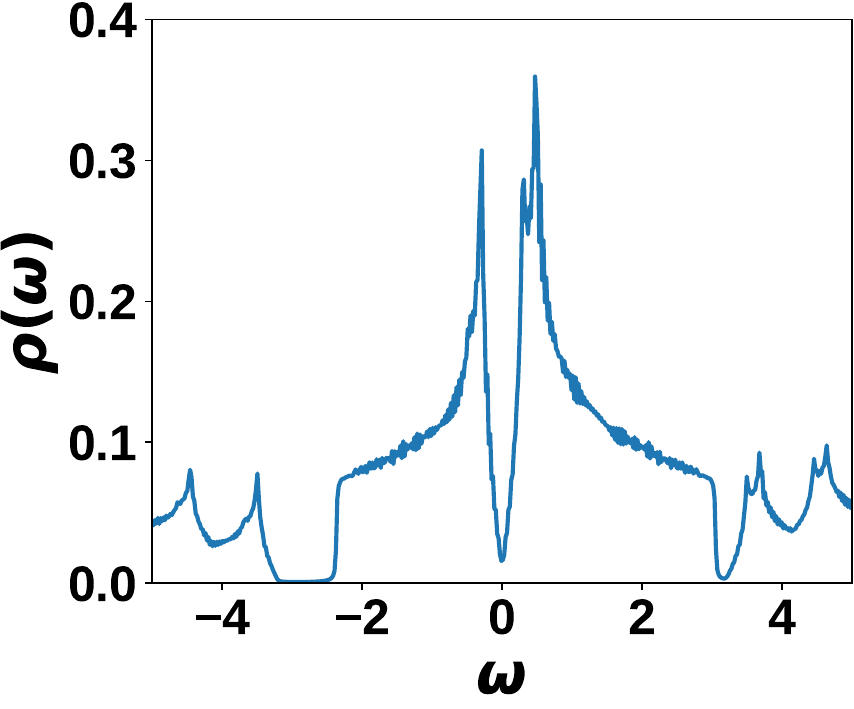}
        \put(87,70){\color{red}\small\bfseries (a)}
    \end{overpic}
    
\end{subfigure}
\hfill
\begin{subfigure}[t]{0.23\textwidth}
    \centering
    \begin{overpic}[width=\linewidth]{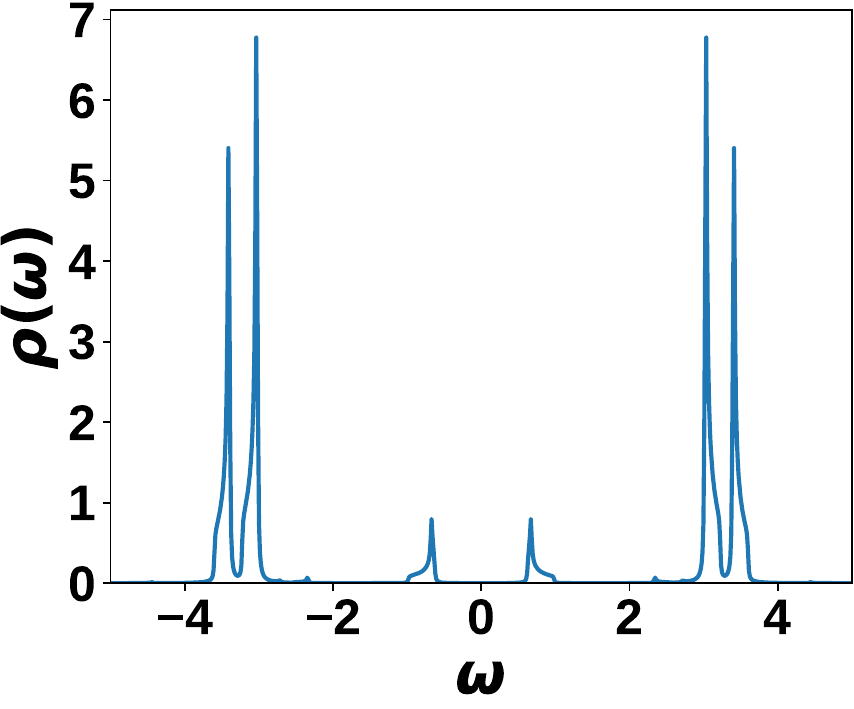}
        \put(87,70){\color{red}\small\bfseries (b)}
    \end{overpic}
    
\end{subfigure}

\caption{\justifying Total DOS for (a) SC phase at \(T = 0.01, \delta = 0.20\)  and (b) AFM + SC + PDW phase at \(T = 0.01\), \(\delta = 0.04\). The parameters correspond to representative points within the respective phases of the phase diagram.}
\end{figure} 

Fig. 11(a) shows the total density of state of SC state exhibits a pronounced V-shaped suppression around the Fermi level, characteristic of a d-wave SC state. Owing to the finite doping level, the spectrum is particle-hole asymmetric. In addition to the coherent features, higher-energy incoherent features arising from the rotor sector are visible. Similar to the AFM phase, the dominant incoherent peaks exhibit a weak pairwise splitting originating from intersite rotor correlations. 

We next examine the total density of states in the coexistence region where AFM, singlet SC, and triplet PDW orders are simultaneously present. As shown in Fig. 11(b), the low-energy spectrum is dominated by the AFM gap, resulting in a pronounced suppression of spectral weight at the Fermi level and the emergence of prominent coherence peaks at finite energies. Although both SC and PDW order parameters are finite in this phase, their signatures are not clearly resolved in the density of states. This behavior can be attributed to the hierarchy of energy scales in the coexistence regime, where the AFM gap is substantially larger than the corresponding SC and PDW gap amplitudes. Consequently, the overall structure of the DOS remains predominantly AFM-like, while the SC and PDW orders provide comparatively small corrections to the quasiparticle spectrum. In addition, a series of higher-energy incoherent peaks is observed, arising from rotor excitations. Similar to the AFM phase, the dominant incoherent peaks exhibit a characteristic pairwise splitting associated with intersite rotor correlations. \\

\begin{figure}[htbp]
\centering


\begin{subfigure}[t]{0.250\textwidth}
    \centering
    \begin{overpic}[width=\linewidth]{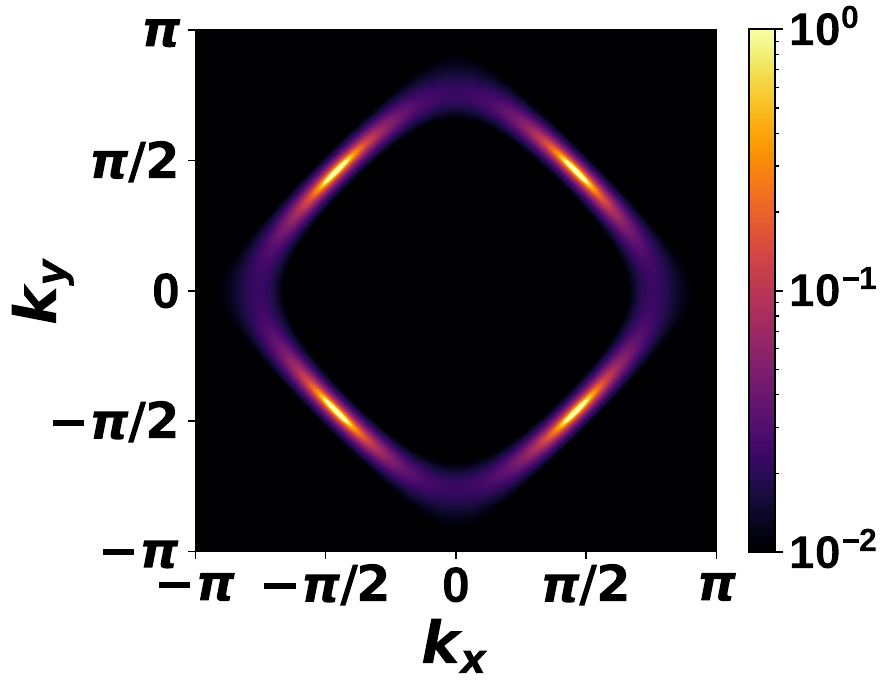}
        \put(68,65){\color{red}\small\bfseries (a)}
    \end{overpic}
\end{subfigure}
\hfill
\begin{subfigure}[t]{0.228\textwidth}
    \centering
    \begin{overpic}[width=\linewidth]{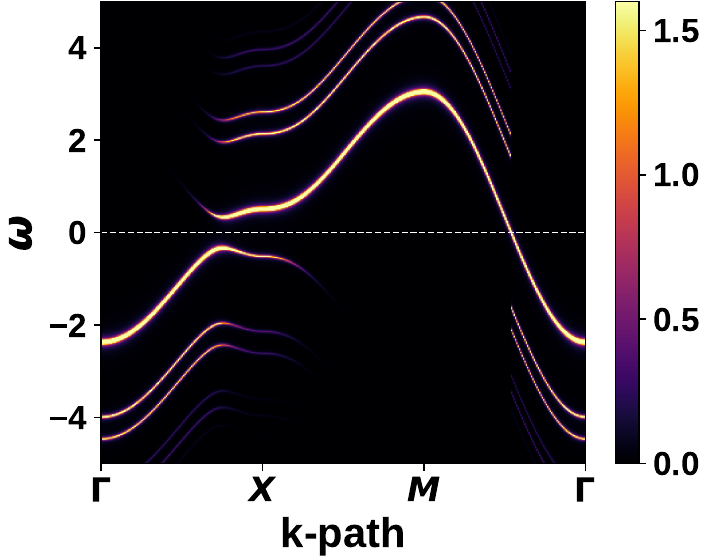}
        \put(70,70){\color{red}\small\bfseries (b)}
    \end{overpic}
    
\end{subfigure}

\caption{\justifying (a) Intensity map of the coherent spectral function \(A(\mathbf{k},\omega=0)\) in the two-dimensional Brillouin zone for the \(d\)-wave SC phase shown in Fig. 10(a) and (b) Momentum-resolved spectral functions \(A(k,\omega)\) along the along the high-symmetry path \(\Gamma\)-\(X\)-\(M\)-\(\Gamma\) for the same SC phase}

\label{fig:SC_PDW_maps}

\end{figure}

To further characterize the SC state, we examine the coherent zero-energy spectral intensity \((A(\mathbf{k},\omega=0))\) in the Brillouin zone. As shown in Fig. 12(a), the spectral weight is concentrated at four isolated points along the Fermi surface, corresponding to the intersections of the Fermi surface with the nodal lines of the \(d_{x^2-y^2}\)-wave superconducting gap. These high-intensity regions identify the nodal quasi-particles, where the superconducting gap vanishes and gapless Bogoliubov excitations persist. Away from the nodal points, the finite \(d\)-wave gap suppresses the low-energy spectral weight, resulting in a strong reduction of \(A(\mathbf{k},\omega=0)\). The emergence of these nodal quasi-particles is consistent with the \(V\)-shaped low-energy density of states and provides direct momentum-space evidence for the \(d\)-wave SC nature of the phase. \\

Fig. - 12(b) shows the total momentum-resolved spectral function \(A(\textbf{k},\omega)\) along the \(\Gamma\)-\(X\)-\(M\)-\(\Gamma\) path in the SC phase. The low-energy spectrum is dominated by a coherent quasiparticle band that crosses the Fermi level along the \(M\)-\(\Gamma\) segment, reflecting the existence of gapless nodal excitations characteristic of a \(d_{x^{2} - y^{2}}\)-wave superconducting state. The absence of a crossing along the \(\Gamma\)-\(X\) and \(X\)-\(M\) segments arises from the finite superconducting gap away from the nodal direction. In addition to the coherent quasi-particle branch, several weaker incoherent bands appear at higher energies due to rotor excitations. A prominent feature of the incoherent spectrum is the pairwise splitting of these bands, which originates from the inter site rotor-correlation term in the rotor Hamiltonian. This term hybridizes local rotor excitations and lifts their degeneracy, resulting in the characteristic split-band structure observed in the spectral function.

 \section{CONCLUSIONS}

In this work, we have investigated the role of strong electronic correlations in the dynamical generation of \(d\)-wave triplet pair-density-wave (PDW) order in a model with strong correlation and described on the 2D square lattice within the slave–rotor cluster mean-field framework. By treating charge and spin degrees of freedom separately, we demonstrated that the interplay between antiferromagnetism (AFM) and \(d\)-wave singlet superconductivity (SC) can naturally induce PDW order without introducing explicit PDW interactions. \\ 

Our analysis of the interaction-strength dependence further highlights the crucial role of strong correlations in stabilizing the PDW state. As \(U\) increases, the expansion of the AFM phase and the concurrent suppression of the uniform SC dome confine superconductivity to an AFM background, thereby enlarging the AFM–SC coexistence regime. This regime provides the necessary environment for the dynamical emergence of PDW order, underscoring that finite-momentum pairing is not an independent instability but is instead driven by the interplay between strong AFM correlations and SC. The stabilization of PDW correlations at intermediate to strong coupling thus points to a correlation-induced mechanism for PDW formation within the slave–rotor framework. \\

The doping and temperature evolution of the charge and pairing sectors reveals a strong interplay between charge coherence, antiferromagnetism, superconductivity, and pair-density-wave order. The rotor condensate \(\phi\), which measures charge coherence, increases with doping and decreases with temperature, reflecting the enhancement of charge fluctuations away from half-filling and the suppression of phase coherence by thermal fluctuations. In contrast, the AFM order parameter is progressively weakened by both doping and temperature, leading to a well-defined magnetic phase boundary in the \(T-\delta\) plane. The SC order parameter exhibits a dome-shaped dependence on doping, reaching its maximum at intermediate doping where magnetic correlations are sufficiently weakened while pairing correlations remain strong. Furthermore, a finite PDW order emerges dynamically only within a restricted region where AFM and singlet SC orders coexist. The PDW phase occupies a much narrower region of the phase diagram and is strongly suppressed by increasing temperature, highlighting its intertwined nature and its dependence on both charge coherence and the coexistence of AFM and SC correlations. \\

Within the slave-rotor mean-field theory, we find that the tetracritical point and the associated PDW phase exist only in the regime \(W_m > W_s\), where AFM and SC coexist. Increasing \(W_m\) suppresses the superconducting dome and ultimately eliminates the AFM–SC phase boundary, while larger \(W_s\) rapidly shrinks the parameter window supporting the coexistence region. \\

Finally, we analyze the spectral signatures associated with the coherent-to-incoherent-metal crossover through the evolution of the total density of states. At low temperatures, a finite rotor condensate \(\phi\) leads to well-defined quasiparticle peaks and a coherent spectral weight at the Fermi level. With increasing temperature, the progressive suppression of the condensate reduces the quasiparticle weight and gives rise to incoherent sidebands, signaling the onset of spectral weight transfer. In the high-temperature regime, where the rotor condensate \(\phi\) vanishes, the spectrum becomes fully incoherent and is characterized by broadened Hubbard-like features with strongly suppressed low-energy weight, indicative of incoherent-metal behavior. Within the AFM+\((\phi\neq0)\) phase, the density of states and momentum-resolved spectral function exhibit a coexistence of coherent and incoherent features separated by an antiferromagnetic gap. In contrast, the AFM+\((\phi=0)\) phase retains the incoherent Hubbard-like excitations while coherent quasiparticle bands are completely absent. A characteristic feature of the incoherent spectrum is the pairwise splitting of the Hubbard-like bands, which originates from the intersite rotor-correlation term in the rotor Hamiltonian. In the superconducting phase, the density of states exhibits the characteristic low-energy behavior of a \(d_{x^{2}-y^{2}}\)-wave superconductor, while the zero-energy spectral intensity reveals the emergence of nodal quasiparticles at the intersections of the Fermi surface and the nodal lines of the superconducting gap. These results demonstrate how charge coherence, rotor excitations, and competing ordered phases collectively shape the low-energy electronic spectrum within the slave-rotor framework.\\

In summary, our results provide a unified picture of how strong electronic correlations govern both ordering phenomena and spectral properties in correlated systems. The slave–rotor framework captures the simultaneous emergence of PDW order, the restructuring of the phase diagram with interaction strength, and the evolution from coherent quasiparticles to incoherent excitations. The identification of distinct regimes characterized by charge coherence, together with the correlation-driven spectral weight redistribution, highlights the central role of charge–spin separation in shaping the low-energy physics. These findings not only clarify the mechanism of dynamical PDW formation but also offer broader insight into the interplay of competing orders and incoherent metallic behavior, which may be relevant to a wide class of strongly correlated materials. \\

\section*{ACKNOWLEDGEMENTS}

We acknowledge National Supercomputing Mission (NSM) for providing computing resources of "PARAMShakti" at IIT Kharagpur, which is implemented by C-DAC and supported by the Ministry of Electronics and Information Technology (MeitY) and Department of Science and Technology (DST), Goverment of India. S.M. also acknowledge the Ministry of Education, Govt. of India for a research fellowship. 

\appendix
\section{SLAVE ROTOR CALCULATION} 
\begin{equation}
\begin{aligned}
H =& -t\sum_{\langle i,j \rangle ,\sigma}
(c_{i,\sigma}^{\dagger}c_{j,\sigma} + h.c.)
+ U\sum_{i} n_{i,\uparrow}n_{i,\downarrow} \\
& - W_{s}\sum_{i}(C_{i}^{\dagger}\langle C_{i} \rangle + h.c.)
- W_{p}\sum_{i}(D_{i}^{\dagger}\langle D_{i}\rangle + h.c.)
\end{aligned}
\end{equation}
In SRMFT, \(c_{i,\sigma} = f_{i,\sigma}e^{-i\theta_i}, c_{i,\sigma}^{\dagger} = f_{i,\sigma}^{\dagger}e^{i\theta_i}\) 
\begin{equation}
{c_{i,\sigma}^{\dagger}c_{j,\sigma}} = f_{i,\sigma}^{\dagger}f_{j,\sigma}e^{i\theta_i}e^{-i\theta_j}
\end{equation}
\begin{equation}
\begin{aligned}
C_{i} &= \sum_{\eta}\zeta(\eta)
\bigl(c_{i,\uparrow}c_{i+\eta,\downarrow}
- c_{i,\downarrow}c_{i+\eta,\uparrow}\bigr) \\
&= \sum_{\eta}\zeta(\eta)
\bigl(f_{i,\uparrow}f_{i+\eta,\downarrow}
- f_{i,\downarrow}f_{i+\eta,\uparrow}\bigr)e^{-i\theta_i}e^{-i\theta_{i+\eta}}\\ 
\end{aligned}
\end{equation}
\begin{equation}
\begin{aligned}
D_{i} &= \sum_{\eta}\zeta(\eta)
\bigl(c_{i,\uparrow}c_{i+\eta,\downarrow}
+ c_{i,\downarrow}c_{i+\eta,\uparrow}\bigr)\\
&= \sum_{\eta}\zeta(\eta)
\bigl(f_{i,\uparrow}f_{i+\eta,\downarrow}
+ f_{i,\downarrow}f_{i+\eta,\uparrow}\bigr)e^{-i\theta_i}e^{-i\theta_{i+\eta}} \\
\end{aligned}
\end{equation}
\begin{equation}
\begin{aligned}
C_{f,i} &= \sum_{\eta}\zeta(\eta)
\bigl(f_{i,\uparrow}f_{i+\eta,\downarrow}
- f_{i,\downarrow}f_{i+\eta,\uparrow}\bigr)\\
D_{f,i} &= \sum_{\eta}\zeta(\eta)
\bigl(f_{i,\uparrow}f_{i+\eta,\downarrow}
- f_{i,\downarrow}f_{i+\eta,\uparrow}\bigr)   \\
&C_{i}^{\dagger}C_i \rightarrow C_{f,i}^{\dagger}C_{f,i}  \:\: D_i^{\dagger}D_i \rightarrow D_{f,i}^{\dagger}D_{f,i}
\end{aligned}
\end{equation}
\begin{equation}
\begin{aligned}
U\sum\limits_{i}^{}&n_{i\uparrow}n_{i,\downarrow}=\frac{U}{4}\sum\limits_i^{}\Big[(n_{i,\uparrow} + n_{i,\downarrow})^{2} - (n_{i,\uparrow} - n_{i,\downarrow})^{2}\Big]\\
&=\frac{U}{4}\sum\limits_{i}n_i^{2} - \frac{U}{4}\sum\limits_{i}\big((n_{i,\uparrow} - n_{i,\downarrow})\langle n_{i,\uparrow} - n_{i,\downarrow}\rangle + h.c. \big) \\
&\text{in slave rotor transformation}\\
&=\frac{U}{4}\sum\limits_{i}(n_i^{\theta})^{2} - \frac{U}{4}\sum\limits_{i}\big((n_{i,\uparrow}^{f} - n_{i,\downarrow}^{f})\langle n_{i,\uparrow}^{f} - n_{i,\downarrow}^{f}\rangle \\
&+ h.c. \big) =\frac{U}{4}\sum\limits_{i}(n_i^{\theta})^{2} - \frac{U}{4}\sum\limits_{i}\big(M_{f,i}^{\dagger}\langle M_{f,i} \rangle + h.c. \big )
\end{aligned}
\end{equation}
we can write the Hamiltonian in terms of spinon and rotor field opeartor,
\begin{equation}
\begin{aligned}
H_{SR} =& -t\sum\limits_{\langle i,j,\sigma \rangle}(f_{i,\sigma}^{\dagger}f_{j,\sigma}e^{i\theta_i}e^{-i\theta_j} + h.c.) + \frac{U}{4}\sum\limits_{i}(n_i^{\theta})^{2}\\
&-\frac{U}{4}\sum\limits_{i}\big(M_{f,i}^{\dagger}\langle M_{f,i} \rangle + h.c. \big ) -  W_{s} \sum\limits_{i} (C_{f,i}^{\dagger} \langle C_{f,i} \rangle  \\
&+ h.c.) -  W_{p}\sum\limits_{i} (D_{f,i}^{\dagger} \langle D_{f,i} \rangle + h.c.)
\end{aligned}
\end{equation}
Now, reduced spinon and rotor Hamiltonians are defined by, 
\begin{equation}
\begin{aligned}
\mathcal{H}_{f} =&  \langle \psi_\theta|H_{SR}|\psi_\theta\rangle = -t\sum\limits_{\langle i,j,\sigma \rangle} (f_{i,\sigma}^{\dagger}f_{j,\sigma}B_{i,j} + h.c.) \\
&+ \frac{U}{4}\sum\limits_{i}\langle(n_{i}^{\theta})^{2}\rangle_{\theta}
-\frac{U}{4}\sum\limits_{i}\big(M_{f,i}^{\dagger}\langle M_{f,i} \rangle + h.c. \big ) \\
&-  W_{s} \sum\limits_{i} (C_{f,i}^{\dagger} \langle C_{f,i} \rangle + h.c.) -  W_{p}\sum\limits_{i} (D_{f,i}^{\dagger} \langle D_{f,i} \rangle \\
&+ h.c.)
\end{aligned}
\end{equation}
\begin{equation}
\begin{aligned}
\mathcal{H}_{\theta} =&  \langle \psi_f|H_{SR}|\psi_f\rangle = -2t\sum\limits_{\langle i,j\rangle} (\chi_{i,j}e^{i\theta_i}e^{-i\theta_j} + h.c.)  \\
&+ \frac{U}{4}\sum\limits_{i}(n_{i}^{\theta})^{2}\rangle
-\frac{U}{4}\sum\limits_{i}\langle\big(M_{f,i}^{\dagger}\langle M_{f,i} \rangle + h.c. \big )\rangle_f \\
&-  W_{s} \sum\limits_{i} \langle(C_{f,i}^{\dagger} \langle C_{f,i} \rangle + h.c.)\rangle_f \\
&-  W_{p}\sum\limits_{i} \langle(D_{f,i}^{\dagger} \langle D_{f,i} \rangle + h.c.)\rangle_f
\end{aligned}
\end{equation}
The ground state energy is \(\langle \Psi_f|\mathcal{H}_f|\Psi_f\rangle = \langle \Psi_\theta|\mathcal{H}_\theta|\Psi_\theta\rangle\). To get this, we must self-consistently solve two coupled Hamiltonians. 
\begin{equation}
\begin{aligned}
{H}_{f} &= -t\sum\limits_{\langle i,j,\sigma \rangle} (f_{i,\sigma}^{\dagger}f_{i,\sigma}B_{i,j} + h.c.) \\
&-\frac{U}{4}\sum\limits_{i}\big(M_{f,i}^{\dagger}\langle M_{f,i} \rangle + h.c. \big ) -  W_{s} \sum\limits_{i} (C_{f,i}^{\dagger} \langle C_{f,i} \rangle \\
&+ h.c.) -  W_{p}\sum\limits_{i} (D_{f,i}^{\dagger} \langle D_{f,i} \rangle + h.c.) - \mu_f\sum\limits_{i,\sigma}n_{i,\sigma}^f
\end{aligned}
\end{equation}
\begin{equation}
\begin{aligned}
H_{\theta} =& -2t\sum\limits_{\langle i,j\rangle} (\chi_{i,j}e^{i\theta_i}e^{-i\theta_j} + h.c.) + \frac{U}{4}\sum\limits_{i}(n_{i}^{\theta})^{2}\rangle \\
&-\mu_\theta\sum\limits_{i}n_{i}^{\theta}
\end{aligned}
\end{equation}
\noindent The chemical potentials \(\mu_f\) and \(\mu_\theta\) impose the average number constraint \(\sum\limits_{\sigma}\langle n_{i,\sigma}^{f} \rangle = 1 - \delta, \:\: \langle n_{i}^{\theta}\rangle = \delta\). Inserting the order parameter in the spinon Hamiltonian,
\begin{equation}
\begin{aligned}
H_f &= -t\sum\limits_{\langle i,j,\sigma \rangle} (f_{i,\sigma}^{\dagger}f_{i,\sigma}B_{i,j} + h.c.)\\
&-\frac{U}{4} \Delta_m\sum\limits_{i}\big(M_{f,i}^{\dagger}\cos(Q.r_i) + h.c. \big) -  W_{s}\Delta_{sf} \sum\limits_{i} (C_{f,i}^{\dagger} \\
&+ h.c.) - W_{p}\Delta_{pf}\sum\limits_{i} (D_{f,i}^{\dagger}\cos(Q.r_i) + h.c.) - \mu_f\sum\limits_{i,\sigma}n_{i,\sigma}^f
\end{aligned}
\end{equation}
In the Nambu spinor basis, we define \\
\(\Psi_{k}^{\dagger} = \{f_{k,\uparrow}^{\dagger}, f_{-k,\downarrow}, f_{k+Q, \uparrow}^{\dagger}, f_{-k-Q,\downarrow}\} \) 
\begin{equation}
H_k = 
\begin{pmatrix}
    \epsilon_k & W_s\Delta_{sf}g(k) & - U\Delta_m & W_p\Delta_{pf}g(k) \\
    W_s\Delta_{sf}g & -\epsilon_k & -W_p\Delta_{pf}g(k) & -U\Delta_m \\
    -U\Delta_m & -W_p\Delta_{pf}g(k) & \epsilon_{k+Q} & -W_s \Delta_{sf}g(k) \\
    W_p\Delta_{pf}g(k) & -U\Delta_m & -W_s\Delta_{sf}g(k) & -\epsilon_{k+Q} \\
\end{pmatrix} \\
\end{equation}
Digonalising the Hamiltonian, the eigenvalues are \(E_{\alpha=\pm}\). 
\begin{equation}
\begin{aligned}
 E_{\alpha} =& \big((\epsilon_k^{2} + \epsilon_{k+Q}^{2})/2 + (U\Delta_m)^{2} + (W_s\Delta_{sf}g(k))^{2} \\
 &+ (W_p\Delta_{pf}g(k))^{2} +(\alpha/2) f\big)^{0.5}  
\end{aligned}
\end{equation}
\begin{equation}
\begin{aligned}
f =& \big((\epsilon_k^{2}-\epsilon_{k+Q}^{2})^{2}+4(U\Delta_m)^{2}(\epsilon_k + \epsilon_{k+Q})^{2} + 4(W_p\Delta_{pf}g(k))^{2}\\
&(\epsilon_k-\epsilon_{k+Q})^{2} + 16(U\Delta_m)(W_s\Delta_{sf}g(k))(W_p\Delta_{pf}g(k))\\
&(\epsilon_k + \epsilon_{k+Q}) + 16(W_s\Delta_{sf}g(k))^{2}(W_p\Delta_{pf}g(k))^{2}\big)^{0.5}
\end{aligned}
\end{equation}
The resulting free energy is
\begin{equation}
\begin{aligned}
F = & -2kT\sum\limits_{k}^{}\sum\limits_{\alpha=\pm}\log\big(2\cosh(\frac{\beta E_{\alpha}(k)}{2})\big) + E_0
\end{aligned}    
\end{equation}

\noindent where, \(E_0\) comes from mean-field decoupling.
\section{ROTOR HAMILTONIAN AND THERMAL AVERAGES} 
The two-site Rotor Hamiltonian \(H_\theta = -2t\chi(e^{i\theta_1}e^{-i\theta_2} + h.c.) - 6t\chi\phi(e^{i\theta_1}+e^{i\theta_2} + h.c.) + \frac{U}{4}(n_1^{\theta})^{2} + \frac{U}{4}(n_2^{\theta})^{2} - \mu_\theta(n_1^{\theta} + n_2^{\theta})\) \\
In the \(|n_1^{\theta}, n_2^{\theta} \rangle\) basis, we calculate the matrix element. 
\begin{equation}
\begin{aligned}
&\langle n_1^{\theta},n_2^{\theta}|H_\theta|n_1^{\theta'},n_2^{\theta'}\rangle \\
=& -2t\chi\langle n_1^{\theta},n_2^{\theta}|e^{i\theta_1}e^{-i\theta_2}+h.c.|n_1^{\theta'},n_2^{\theta'}\rangle \\
&- 6t\chi\phi\langle n_1^{\theta},n_2^{\theta}|e^{i\theta_1}+e^{i\theta_2} + h.c.|n_1^{\theta'},n_2^{\theta'}\rangle  \\
&+ \frac{U}{4}\langle n_1^{\theta},n_2^{\theta}|(n_1^{\theta})^{2} + (n_2^{\theta})^{2}|n_1^{\theta'},n_2^{\theta'}\rangle \\
&- \mu_\theta \langle n_1^{\theta},n_2^{\theta}| n_1^{\theta} + n_2^{\theta} |n_1^{\theta'},n_2^{\theta'}\rangle \\
=& -2t\chi(\delta_{n_1^{\theta},n_2^{\theta'}+1}\delta_{n_2^{\theta},n_2^{\theta'}-1} + \delta_{n_1^{\theta},n_2^{\theta'}-1}\delta_{n_2^{\theta},n_2^{\theta'}+1})\\
&-6t\chi\phi(\delta_{n_1^{\theta},n_1^{\theta'}+1}\delta_{n_2^{\theta},n_2^{\theta'}}  +\delta_{n_1^{\theta},n_1^{\theta'}}\delta_{n_2^{\theta},n_2^{\theta'}+1}+\delta_{n_1^{\theta},n_1^{\theta'}-1}\delta_{n_2^{\theta},n_2^{\theta'}}\\
&+\delta_{n_1^{\theta},n_1^{\theta'}}\delta_{n_2^{\theta},n_2^{\theta'}-1}) +  \big(\frac{U}{4}\big((n_1^{\theta})^{2} + (n_2^{\theta})^{2}\big) - \mu_\theta(n_1^{\theta} + n_2^{\theta})\big)\\
&\delta_{n_1^{\theta},n_1^{\theta'}}\delta_{n_2^{\theta},n_2^{\theta'}}
\end{aligned}
\end{equation}
Thermal evolution of \(\phi\) and \(B\),
\begin{equation}
\langle \phi \rangle_{\text{avg}} = \frac{\text{Tr}(\phi e^{-\beta H_\theta})}{\text{Tr}(e^{-\beta H_\theta})}, \:\: \langle B \rangle_{\text{avg}} = \frac{\text{Tr}(B e^{-\beta H_\theta})}{\text{Tr}(e^{-\beta H_\theta})}
\end{equation}
\section{CALCULATION OF SPECTRAL FUNCTION}
The electronic Green's function is an convolution of spinon and rotor Green's function. 
\begin{equation}
\begin{aligned}
&G_c(\textbf{k}, i\omega_n) = \frac{1}{\beta N}\sum\limits_{q,m}G_f(\textbf{k - q}, i\omega_n - i\nu_m)G_{\theta}(q, i\nu_m)\\
&\text{where}, i\omega_n \:\text{and}\: i\nu_m \: \text{are fermionic and bosonic  Matsubara
}\\
&\text{frequencies respectively.}
\end{aligned}
\end{equation}
\begin{equation}
    G_\theta(\textbf{q},i\nu_m) = \beta|\phi|^{2}\delta_{q,0}\delta_{m,0} + G_{\theta}^{\mathrm{inc}}(\textbf{q},i\nu_m)
\end{equation}
The first term represents a zero-momentum, zero-frequency coherent rotor condensate. It does not describe a finite-energy transition between different rotor eigenstates.
In contrast, \(G_{\theta}^{\text{inc}}\) describes the finite-energy charge fluctuations and contains the transitions between different rotor eigenstates. 
\begin{equation}
\begin{aligned}
G_c(\textbf{k}, &i\omega_n) = \frac{1}{\beta N}\sum\limits_{q,m}G_f(\textbf{k - q}, i\omega_n - i\nu_m)\big(\beta|\phi|^{2}\delta_{q,0}\delta_{m,0} \\
&+ G_{\theta}^{\mathrm{inc}}(\textbf{q},i\nu_m)\big)
=|\phi|^{2}G_f(\textbf{k},i\omega_n) + G_c^{\text{inc}}(\textbf{k},i\omega_n)
\end{aligned}
\end{equation}
\begin{equation}
\begin{aligned}
G_c^{\text{coh}}(\textbf{k},i\omega_n) = ZG_f(\textbf{k},i\omega_n)= Z\Bigg[\frac{u_k^{2}}{i\omega_n -E_k} + \frac{v_k^{2}}{i\omega_n + E_k}\Bigg]
\end{aligned}
\end{equation}
After analytic continuation, \(i\omega_n \longrightarrow \omega + i0^{+}\)
\begin{equation}
\begin{aligned}
A_{\text{coh}}(\textbf{k},\omega)=Z  \Bigg[ u_k^{2}\delta(\omega - E_{\mathbf{k}}) + v_k^{2}\delta(\omega + E_{\mathbf{k}})\Bigg]
\end{aligned}
\end{equation}
Now for the incoherent part,
\begin{equation}
G_c^{\text{inc}}(\textbf{k},i\omega_n) = \frac{1}{\beta N}\sum\limits_{q,m}G_f(\textbf{k - q}, i\omega_n - i\nu_m)G_{\theta}^{\text{inc}}(\textbf{q},i\nu_m)
\end{equation}
Using Lehmann representation for the rotor sector, 
\begin{equation}
G_\theta(\textbf{q}, i\nu_n) = \sum\limits_{m\neq n} \frac{P_m - P_n}{i\omega_n - w_{mn}} \langle \psi_m|e^{-i\theta}|\psi_n\rangle|^{2}
\end{equation}
\(P_m = e^{-\beta w_m}/Z_{\text{par}}, \:\:P_n = e^{-\beta w_n}/Z_{\text{par}}, \: w_{mn} = w_m - w_n\)
\begin{equation}
G_f(\textbf{k},i\omega_n - i\nu_m) = \frac{u_k^{2}}{i\omega_n - i\nu_m - E_k} + \frac{v_k^{2}}{i\omega_n + i\nu_m - E_k}   
\end{equation}
Using Matsubara summation, and putting $G_f$, the incoherent spectral function is obtained.

\bibliographystyle{apsrev4-2}
\bibliography{mybib}

@article{RevModPhys.87.457,
  title = {Colloquium: Theory of intertwined orders in high temperature superconductors},
  author = {Fradkin, Eduardo and Kivelson, Steven A. and Tranquada, John M.},
  journal = {Rev. Mod. Phys.},
  volume = {87},
  issue = {2},
  pages = {457--482},
  numpages = {26},
  year = {2015},
  month = {May},
  publisher = {American Physical Society},
  doi = {10.1103/RevModPhys.87.457},
  url = {https://link.aps.org/doi/10.1103/RevModPhys.87.457}
}

@article{Tu2016,
  title = {Genesis of Charge Orders in High-Temperature Superconductors},
  author = {Tu, Wei-Lin and Lee, Ting-Kuo},
  journal = {Sci. Rep.},
  volume = {6},
  issue = {1},
  pages = {18675},
  numpages = {0},
  year = {2016},
  month = {Jan},
  publisher = {Nature Publishing Group},
  doi = {10.1038/srep18675},
  url = {https://doi.org/10.1038/srep18675}
}

@article{Hayden2024,
  title = {Charge Correlations in Cuprate Superconductors},
  author = {Hayden, Stephen M. and Tranquada, John M.},
  journal = {Annu. Rev. Condens. Matter Phys.},
  volume = {15},
  issue = {1},
  pages = {215--235},
  numpages = {21},
  year = {2024},
  month = {Mar},
  doi = {10.1146/annurev-conmatphys-032922-094430},
  url = {https://doi.org/10.1146/annurev-conmatphys-032922-094430}
}

@article{PhysRev.135.A550,
  title = {Superconductivity in a Strong Spin-Exchange Field},
  author = {Fulde, Peter and Ferrell, Richard A.},
  journal = {Phys. Rev.},
  volume = {135},
  issue = {3A},
  pages = {A550--A563},
  numpages = {0},
  year = {1964},
  month = {Aug},
  publisher = {American Physical Society},
  doi = {10.1103/PhysRev.135.A550},
  url = {https://link.aps.org/doi/10.1103/PhysRev.135.A550}
}

@article{Larkin1965,
  title = {Nonuniform State of Superconductors},
  author = {Larkin, A. I. and Ovchinnikov, Yu. N.},
  journal = {Sov. Phys. JETP},
  volume = {20},
  issue = {3},
  pages = {762--769},
  numpages = {8},
  year = {1965},
  month = {Mar},
  url = {http://www.jetp.ac.ru/cgi-bin/e/index/e/20/3/p762?a=list}
}

@article{PhysRevLett.88.117001,
  title = {Stripe States with Spatially Oscillating $\mathit{d}$-Wave Superconductivity in the Two-Dimensional $\mathit{t}\ensuremath{-}{\mathit{t}}^{\ensuremath{'}}\ensuremath{-}\mathit{J}$ Model},
  author = {Himeda, A. and Kato, T. and Ogata, M.},
  journal = {Phys. Rev. Lett.},
  volume = {88},
  issue = {11},
  pages = {117001},
  numpages = {4},
  year = {2002},
  month = {Feb},
  publisher = {American Physical Society},
  doi = {10.1103/PhysRevLett.88.117001},
  url = {https://link.aps.org/doi/10.1103/PhysRevLett.88.117001}
}

@article{PhysRevLett.99.127003,
  title = {Dynamical Layer Decoupling in a Stripe-Ordered High-${T}_{c}$ Superconductor},
  author = {Berg, E. and Fradkin, E. and Kim, E.-A. and Kivelson, S. A. and Oganesyan, V. and Tranquada, J. M. and Zhang, S. C.},
  journal = {Phys. Rev. Lett.},
  volume = {99},
  issue = {12},
  pages = {127003},
  numpages = {4},
  year = {2007},
  month = {Sep},
  publisher = {American Physical Society},
  doi = {10.1103/PhysRevLett.99.127003},
  url = {https://link.aps.org/doi/10.1103/PhysRevLett.99.127003}
}

@article{Agterberg2008,
  title = {Dislocations and Vortices in Pair-Density-Wave Superconductors},
  author = {Agterberg, D. F. and Tsunetsugu, H.},
  journal = {Nat. Phys.},
  volume = {4},
  issue = {8},
  pages = {639--642},
  numpages = {4},
  year = {2008},
  month = {Aug},
  publisher = {Nature Publishing Group},
  doi = {10.1038/nphys1017},
  url = {https://doi.org/10.1038/nphys1017}
}

@article{Agterberg2020,
  title = {The Physics of Pair-Density Waves: Cuprate Superconductors and Beyond},
  author = {Agterberg, Daniel F. and Davis, J. C. S\'eamus and Edkins, Stephen D. and Fradkin, Eduardo and Van Harlingen, Dale J. and Kivelson, Steven A. and Lee, Patrick A. and Radzihovsky, Leo and Tranquada, John M. and Wang, Yuxuan},
  journal = {Annu. Rev. Condens. Matter Phys.},
  volume = {11},
  issue = {1},
  pages = {231--270},
  numpages = {40},
  year = {2020},
  month = {Mar},
  doi = {10.1146/annurev-conmatphys-031119-050711},
  url = {https://doi.org/10.1146/annurev-conmatphys-031119-050711}
}

@article{PhysRevB.79.064515,
  title = {Theory of the striped superconductor},
  author = {Berg, Erez and Fradkin, Eduardo and Kivelson, Steven A.},
  journal = {Phys. Rev. B},
  volume = {79},
  issue = {6},
  pages = {064515},
  numpages = {15},
  year = {2009},
  month = {Feb},
  publisher = {American Physical Society},
  doi = {10.1103/PhysRevB.79.064515},
  url = {https://link.aps.org/doi/10.1103/PhysRevB.79.064515}
}

@article{Yang2009,
  title = {Nature of Stripes in the Generalized $t$--$J$ Model Applied to the Cuprate Superconductors},
  author = {Yang, Kai-Yu and Chen, Wei-Qiang and Rice, Thomas M. and Sigrist, Manfred and Zhang, Fu-Chun},
  journal = {New J. Phys.},
  volume = {11},
  issue = {5},
  pages = {055053},
  numpages = {18},
  year = {2009},
  month = {May},
  doi = {10.1088/1367-2630/11/5/055053},
  url = {https://doi.org/10.1088/1367-2630/11/5/055053}
}

@article{PhysRevB.76.140505,
  title = {Unidirectional $d$-wave superconducting domains in the two-dimensional $t\text{\ensuremath{-}}J$ model},
  author = {Raczkowski, Marcin and Capello, Manuela and Poilblanc, Didier and Fr\'esard, Raymond and Ole\ifmmode \acute{s}\else \'{s}\fi{}, Andrzej M.},
  journal = {Phys. Rev. B},
  volume = {76},
  issue = {14},
  pages = {140505(R)},
  numpages = {4},
  year = {2007},
  month = {Oct},
  publisher = {American Physical Society},
  doi = {10.1103/PhysRevB.76.140505},
  url = {https://link.aps.org/doi/10.1103/PhysRevB.76.140505}
}

@article{PhysRevB.38.4596,
  title = {Superconducting properties of ${\mathrm{La}}_{2\mathrm{\ensuremath{-}}\mathrm{x}}$${\mathrm{Ba}}_{\mathrm{x}}$${\mathrm{CuO}}_{4}$},
  author = {Moodenbaugh, A. R. and Xu, Youwen and Suenaga, M. and Folkerts, T. J. and Shelton, R. N.},
  journal = {Phys. Rev. B},
  volume = {38},
  issue = {7},
  pages = {4596--4600},
  numpages = {0},
  year = {1988},
  month = {Sep},
  publisher = {American Physical Society},
  doi = {10.1103/PhysRevB.38.4596},
  url = {https://link.aps.org/doi/10.1103/PhysRevB.38.4596}
}

@article{PhysRevLett.62.2751,
  title = {Structural phase transformations and superconductivity in ${\mathrm{La}}_{2\mathrm{\ensuremath{-}}\mathrm{x}}$${\mathrm{Ba}}_{\mathrm{x}}$${\mathrm{CuO}}_{4}$},
  author = {Axe, J. D. and Moudden, A. H. and Hohlwein, D. and Cox, D. E. and Mohanty, K. M. and Moodenbaugh, A. R. and Xu, Youwen},
  journal = {Phys. Rev. Lett.},
  volume = {62},
  issue = {23},
  pages = {2751--2754},
  numpages = {0},
  year = {1989},
  month = {Jun},
  publisher = {American Physical Society},
  doi = {10.1103/PhysRevLett.62.2751},
  url = {https://link.aps.org/doi/10.1103/PhysRevLett.62.2751}
}

@article{Axe1994,
  title = {Structural Instabilities in Lanthanum Cuprate Superconductors},
  author = {Axe, J. D. and Crawford, M. K.},
  journal = {J. Low Temp. Phys.},
  volume = {95},
  issue = {1-2},
  pages = {271--284},
  numpages = {14},
  year = {1994},
  month = {Apr},
  doi = {10.1007/BF00754035},
  url = {https://doi.org/10.1007/BF00754035}
}

@article{PhysRevB.78.174529,
  title = {Evidence for unusual superconducting correlations coexisting with stripe order in ${\text{La}}_{1.875}{\text{Ba}}_{0.125}{\text{CuO}}_{4}$},
  author = {Tranquada, J. M. and Gu, G. D. and H\"ucker, M. and Jie, Q. and Kang, H.-J. and Klingeler, R. and Li, Q. and Tristan, N. and Wen, J. S. and Xu, G. Y. and Xu, Z. J. and Zhou, J. and v. Zimmermann, M.},
  journal = {Phys. Rev. B},
  volume = {78},
  issue = {17},
  pages = {174529},
  numpages = {13},
  year = {2008},
  month = {Nov},
  publisher = {American Physical Society},
  doi = {10.1103/PhysRevB.78.174529},
  url = {https://link.aps.org/doi/10.1103/PhysRevB.78.174529}
}

@article{Hamidian2016,
  title = {Detection of a Cooper-Pair Density Wave in Bi$_2$Sr$_2$CaCu$_2$O$_{8+x}$},
  author = {Hamidian, M. H. and Edkins, Stephen D. and Joo, Sang Hyun and Kostin, A. and Eisaki, H. and Uchida, S. and Lawler, M. J. and Kim, E.-A. and Mackenzie, A. P. and Fujita, K. and others},
  journal = {Nature},
  volume = {532},
  issue = {7599},
  pages = {343--347},
  numpages = {5},
  year = {2016},
  month = {Apr},
  publisher = {Nature Publishing Group},
  doi = {10.1038/nature17411},
  url = {https://doi.org/10.1038/nature17411}
}

@article{PhysRevX.4.031017,
  title = {Amperean Pairing and the Pseudogap Phase of Cuprate Superconductors},
  author = {Lee, Patrick A.},
  journal = {Phys. Rev. X},
  volume = {4},
  issue = {3},
  pages = {031017},
  numpages = {13},
  year = {2014},
  month = {Jul},
  publisher = {American Physical Society},
  doi = {10.1103/PhysRevX.4.031017},
  url = {https://link.aps.org/doi/10.1103/PhysRevX.4.031017}
}

@article{PhysRevLett.98.067006,
  title = {Amperean Pairing Instability in the U(1) Spin Liquid State with Fermi Surface and Application to $\ensuremath{\kappa}\mathrm{\text{\ensuremath{-}}}(\mathrm{BEDT}\mathrm{\text{\ensuremath{-}}}\mathrm{TTF}{)}_{2}{\mathrm{Cu}}_{2}(\mathrm{CN}{)}_{3}$},
  author = {Lee, Sung-Sik and Lee, Patrick A. and Senthil, T.},
  journal = {Phys. Rev. Lett.},
  volume = {98},
  issue = {6},
  pages = {067006},
  numpages = {4},
  year = {2007},
  month = {Feb},
  publisher = {American Physical Society},
  doi = {10.1103/PhysRevLett.98.067006},
  url = {https://link.aps.org/doi/10.1103/PhysRevLett.98.067006}
}

@article{PhysRevLett.105.146403,
  title = {Pair-Density-Wave Correlations in the Kondo-Heisenberg Model},
  author = {Berg, Erez and Fradkin, Eduardo and Kivelson, Steven A.},
  journal = {Phys. Rev. Lett.},
  volume = {105},
  issue = {14},
  pages = {146403},
  numpages = {4},
  year = {2010},
  month = {Sep},
  publisher = {American Physical Society},
  doi = {10.1103/PhysRevLett.105.146403},
  url = {https://link.aps.org/doi/10.1103/PhysRevLett.105.146403}
}

@article{PhysRevB.49.4261,
  title = {Spin-fluctuation-induced superconductivity and normal-state properties of ${\mathrm{YBa}}_{2}$${\mathrm{Cu}}_{3}$${\mathrm{O}}_{7}$},
  author = {Monthoux, P. and Pines, D.},
  journal = {Phys. Rev. B},
  volume = {49},
  issue = {6},
  pages = {4261--4278},
  numpages = {0},
  year = {1994},
  month = {Feb},
  publisher = {American Physical Society},
  doi = {10.1103/PhysRevB.49.4261},
  url = {https://link.aps.org/doi/10.1103/PhysRevB.49.4261}
}

@article{PhysRevB.66.165111,
  title = {Quantum impurity solvers using a slave rotor representation},
  author = {Florens, Serge and Georges, Antoine},
  journal = {Phys. Rev. B},
  volume = {66},
  issue = {16},
  pages = {165111},
  numpages = {16},
  year = {2002},
  month = {Oct},
  publisher = {American Physical Society},
  doi = {10.1103/PhysRevB.66.165111},
  url = {https://link.aps.org/doi/10.1103/PhysRevB.66.165111}
}

@article{PhysRevB.70.035114,
  title = {Slave-rotor mean-field theories of strongly correlated systems and the Mott transition in finite dimensions},
  author = {Florens, Serge and Georges, Antoine},
  journal = {Phys. Rev. B},
  volume = {70},
  issue = {3},
  pages = {035114},
  numpages = {15},
  year = {2004},
  month = {Jul},
  publisher = {American Physical Society},
  doi = {10.1103/PhysRevB.70.035114},
  url = {https://link.aps.org/doi/10.1103/PhysRevB.70.035114}
}

@article{PhysRevB.83.134515,
  title = {Magnetism and Mott transition: A slave-rotor study},
  author = {Ko, Wing-Ho and Lee, Patrick A.},
  journal = {Phys. Rev. B},
  volume = {83},
  issue = {13},
  pages = {134515},
  numpages = {6},
  year = {2011},
  month = {Apr},
  publisher = {American Physical Society},
  doi = {10.1103/PhysRevB.83.134515},
  url = {https://link.aps.org/doi/10.1103/PhysRevB.83.134515}
}

@article{PhysRevB.76.195101,
  title = {Self-consistent slave rotor mean-field theory for strongly correlated systems},
  author = {Zhao, E. and Paramekanti, A.},
  journal = {Phys. Rev. B},
  volume = {76},
  issue = {19},
  pages = {195101},
  numpages = {15},
  year = {2007},
  month = {Nov},
  publisher = {American Physical Society},
  doi = {10.1103/PhysRevB.76.195101},
  url = {https://link.aps.org/doi/10.1103/PhysRevB.76.195101}
}

@article{PhysRevB.75.245105,
  title = {How to control pairing fluctuations: SU(2) slave-rotor gauge theory of the Hubbard model},
  author = {Kim, Ki-Seok},
  journal = {Phys. Rev. B},
  volume = {75},
  issue = {24},
  pages = {245105},
  numpages = {16},
  year = {2007},
  month = {Jun},
  publisher = {American Physical Society},
  doi = {10.1103/PhysRevB.75.245105},
  url = {https://link.aps.org/doi/10.1103/PhysRevB.75.245105}
}

@article{Zhang1997,
  title = {A Unified Theory Based on ${SO}(5)$ Symmetry of Superconductivity and Antiferromagnetism},
  author = {Zhang, Shou-Cheng},
  journal = {Science},
  volume = {275},
  issue = {5303},
  pages = {1089--1096},
  numpages = {8},
  year = {1997},
  month = {Feb},
  doi = {10.1126/science.275.5303.1089},
  url = {https://doi.org/10.1126/science.275.5303.1089}
}

@article{Psaltakis1983,
  title = {Superconductivity and Spin-Density Waves: Organic Superconductors},
  author = {Psaltakis, G. C. and Fenton, E. W.},
  journal = {J. Phys. C: Solid State Phys.},
  volume = {16},
  number = {20},
  pages = {3913--3927},
  year = {1983},
  month = {Jul},
  doi = {10.1088/0022-3719/16/20/015},
  url = {https://doi.org/10.1088/0022-3719/16/20/015}
}

@article{PhysRevB.62.9083,
  title = {Mean-field study of the interplay between antiferromagnetism and d-wave superconductivity},
  author = {Kyung, Bumsoo},
  journal = {Phys. Rev. B},
  volume = {62},
  issue = {13},
  pages = {9083--9088},
  numpages = {0},
  year = {2000},
  month = {Oct},
  publisher = {American Physical Society},
  doi = {10.1103/PhysRevB.62.9083},
  url = {https://link.aps.org/doi/10.1103/PhysRevB.62.9083}
}

@article{Maitra2001,
  title = {Antiferromagnetism and Superconductivity in a Model with Extended Pairing Interactions},
  author = {Maitra, Tulika and Beck, H. and Taraphder, A.},
  journal = {Eur. Phys. J. B},
  volume = {21},
  issue = {4},
  pages = {527--533},
  numpages = {7},
  year = {2001},
  month = {Jun},
  doi = {10.1007/PL00011134},
  url = {https://doi.org/10.1007/PL00011134}
}

@article{PhysRevB.89.165126,
  title = {Pair-density-wave superconducting states and electronic liquid-crystal phases},
  author = {Soto-Garrido, Rodrigo and Fradkin, Eduardo},
  journal = {Phys. Rev. B},
  volume = {89},
  issue = {16},
  pages = {165126},
  numpages = {19},
  year = {2014},
  month = {Apr},
  publisher = {American Physical Society},
  doi = {10.1103/PhysRevB.89.165126},
  url = {https://link.aps.org/doi/10.1103/PhysRevB.89.165126}
}

@article{Georges1996,
  title = {Dynamical Mean-Field Theory of Strongly Correlated Fermion Systems and the Limit of Infinite Dimensions},
  author = {Georges, Antoine and Kotliar, Gabriel and Krauth, Werner and Rozenberg, Marcelo J.},
  journal = {Rev. Mod. Phys.},
  volume = {68},
  issue = {1},
  pages = {13--125},
  numpages = {113},
  year = {1996},
  month = {Jan},
  publisher = {American Physical Society},
  doi = {10.1103/RevModPhys.68.13},
  url = {https://link.aps.org/doi/10.1103/RevModPhys.68.13}
}

@article{PhysRevB.91.075124,
  title = {Absence of a quantum limit to charge diffusion in bad metals},
  author = {Pakhira, Nandan and McKenzie, Ross H.},
  journal = {Phys. Rev. B},
  volume = {91},
  issue = {7},
  pages = {075124},
  numpages = {10},
  year = {2015},
  month = {Feb},
  publisher = {American Physical Society},
  doi = {10.1103/PhysRevB.91.075124},
  url = {https://link.aps.org/doi/10.1103/PhysRevB.91.075124}
}

@article{RevModPhys.68.13,
  title = {Dynamical mean-field theory of strongly correlated fermion systems and the limit of infinite dimensions},
  author = {Georges, Antoine and Kotliar, Gabriel and Krauth, Werner and Rozenberg, Marcelo J.},
  journal = {Rev. Mod. Phys.},
  volume = {68},
  issue = {1},
  pages = {13--125},
  numpages = {0},
  year = {1996},
  month = {Jan},
  publisher = {American Physical Society},
  doi = {10.1103/RevModPhys.68.13},
  url = {https://link.aps.org/doi/10.1103/RevModPhys.68.13}
}

@article{PhysRevLett.32.1350,
  title = {Spin Flop, Supersolids, and Bicritical and Tetracritical Points},
  author = {Fisher, Michael E. and Nelson, David R.},
  journal = {Phys. Rev. Lett.},
  volume = {32},
  issue = {24},
  pages = {1350--1353},
  numpages = {0},
  year = {1974},
  month = {Jun},
  publisher = {American Physical Society},
  doi = {10.1103/PhysRevLett.32.1350},
  url = {https://link.aps.org/doi/10.1103/PhysRevLett.32.1350}
}

@article{Acharya2016,
  title = {Feasibility of a Metamagnetic Transition in Correlated Systems},
  author = {Acharya, Swagata and Medhi, Amal and Vidhyadhiraja, N. S. and Taraphder, A.},
  journal = {J. Phys.: Condens. Matter},
  volume = {28},
  issue = {11},
  pages = {116001},
  numpages = {10},
  year = {2016},
  month = {Mar},
  doi = {10.1088/0953-8984/28/11/116001},
  url = {https://doi.org/10.1088/0953-8984/28/11/116001}
}
 
\end{document}